\documentclass[preprint,12pt]{elsarticle}
\usepackage{amssymb}
\usepackage{doi}
\usepackage{hyperref}
\usepackage{amsmath}
\usepackage{graphicx} 
\usepackage{subcaption}
\usepackage{siunitx}
\usepackage{setspace}
\usepackage{booktabs}
\author[1]{Wei Wen}
\fnmark[1]
\ead{wenwei@nuaa.edu.cn}
\address[1]{College of Energy and Power Engineering, Nanjing University of Aeronautics
and Astronautics, Nanjing 210016, China}
\author[1]{Wenkai Qi}
\fnmark[2]
\ead{qwkai@nuaa.edu.cn}
\author[1]{Weidong Wen}
\fnmark[3]

\nonumnote{E-mail addresses: wenwei@nuaa.edu.cn (Wei Wen), qwkai@nuaa.edu.cn (Wenkai Qi).}

\begin{document}
\begin{frontmatter}
\title{Precise Computation of Forced Response Backbone Curves of Frictional Structures Using Analytical Hessian Tensor of Contact Elements}
\begin{abstract}
Predicting the forced vibration response of nonlinear mechanical systems with friction is critical for engineering applications. Accurately determining the backbone curve of resonance peaks is pivotal for the design of friction devices. However, the prediction of these curves is computationally challenging owing to the nonconservative and nonsmooth nature of friction nonlinearity. Although techniques such as damped nonlinear normal modes (dNNMs) and phase resonance methods have been applied, they often suffer from convergence issues, and their computational accuracy is compromised under certain conditions.
This study proposes a novel method for computing the forced response backbone curves of structures with frictional contact interfaces. The method accurately tracks the backbone curve through a parameter continuation scheme, formulated via Lagrange multipliers and accelerated by incorporating a derived analytical Hessian Tensor of contact elements. This approach yields highly accurate numerical results and enables numerical singularities on the curve to be identified and robustly traversed. The proposed method is validated using an Euler–Bernoulli beam finite-element model and a lumped-parameter blade–damper–blade model. The results demonstrate superior accuracy compared to conventional dNNMs and phase resonance methods, particularly in cases involving either high structural damping or strong frictional damping. This work provides a robust computational tool and presents a detailed comparative analysis that clarifies the applicability and limitations of the proposed and conventional methods.
\end{abstract}
\end{frontmatter}
\section{Introduction}
\label{sec:intro}
\subsection{Friction models} 
The friction between contact interfaces is one of the primary sources of nonlinearity in mechanical systems and significantly impacts their dynamic characteristics. In some cases, structural engineers need to mitigate the negative effects of friction, whereas in others, its damping properties must be leveraged to suppress structural vibrations, as observed in underplatform dampers in turbomachinery. In either case, an effective assessment of friction is essential, which has led to extensive studies on friction modelling.

Den \cite{(1)den1931forced} was among the first researchers to study the application of friction models in dynamic analysis, calculating the response of a single-degree of freedom (DOF) system under harmonic excitation using the Coulomb friction model. Subsequently, Iwan \cite{(2)iwan1961dynamic} improved the early Coulomb model by introducing a bilinear hysteretic restoring force to account for the elastic deformation of the contact surface before the critical slip threshold was reached. The Coulomb and Iwan models assume a constant normal load at the contact interface, implying that they do not consider the influence of normal motion on the contact force. To estimate the friction contact more accurately and consider the effect of normal separation on stiffness during large relative displacements, Yang\cite{(3)yang1998stick} developed a three-dimensional friction model that accounts for changes in the contact state at different moments in a periodic motion. This model was widely applied in subsequent studies.

The steady-state response of nonlinear systems with friction interfaces is often calculated using the harmonic balance method (HBM). Nayfeh \cite{(4)nayfeh2024nonlinear} introduced the first-order HBM, which approximates the steady-state periodic solution using only the fundamental harmonic of its Fourier series expansion. Griffin \cite{(5)griffin1980friction} applied a first-order HBM to determine the response of a single-DOF system. To increase the predictive accuracy and account for the influence of superharmonics, the HBM was extended to include a larger number of harmonics \cite{(6)dowell1985multi}\cite{(7)zucca2014nonlinear}. A significant advancement in applying the HBM to systems with friction was the alternating frequency/time-domain (AFT) method proposed by Cameron \cite{(8)cameron1989alternating}, which resolved the challenge of dry friction forces lacking an explicit expression in the frequency domain. The HBM equations are typically solved iteratively using Newton's method, which requires computing the Jacobian matrix of the Fourier coefficients of the nonlinear force with respect to those of the displacement. As the number of DOFs and retained harmonics increases, the computational cost of evaluating the Jacobian via conventional numerical differentiation becomes burdensome. Methods have been developed to calculate the analytical Jacobian, which is the stiffness matrix for the friction contact elements, to accelerate this computation. Petrov and Ewins \cite{(9)petrov2003analytical} derived analytical expressions for the frequency-domain contact forces and stiffness matrices by identifying the state transition times within a period and performing Fourier coefficient integration. Borrajo \cite{(10)borrajo2006analytical} applied this method to calculate the vibration response of turbine blades with wedge friction dampers. Panning \cite{(11)siewert2009multiharmonic} solved for the contact forces and stiffness matrices using a predictor--corrector procedure in the time domain and the discrete Fourier transform.

\subsection{Nonlinear resonance backbone curve prediction}
The forced resonance backbone curve is defined as the locus of the points of the maximum response amplitude on the frequency response function (FRF) under a given level of harmonic excitation. For a slightly damped linear system, the response at or near the resonance peak is dominated by the contribution of a single mode. Consequently, the displacement of the system can be accurately approximated by the response of the corresponding modal coordinate, effectively reducing the multi-DOF system to a single-DOF system within that frequency range. Similar to linear systems, introducing nonlinear normal modes (NNMs) is important for predicting nonlinear resonance backbone curves. Rosenberg \cite{(12)rosenberg1962normal} first introduced the concept of NNMs, defining them as vibration in unison with the system. Shaw and Pierre \cite{(13)shaw1991non} extended Rosenberg's definition by introducing the concept of an invariant manifold, making it applicable to damped systems. This is referred to as damped nonlinear normal modes(dNNMs). However, NNMs under this definition face three main computational challenges: the potential nonuniqueness of the parameterising coordinate \cite{(14)blanc2013numerical}, prohibitive cost of computing the entire manifold for large-scale systems \cite{(15)krack2015nonlinear}, and difficulty handling nonsmooth dynamics.

George Haller \cite{(16)haller2016nonlinear} resolved the uniqueness problem by defining an NNM as the smoothest member of an invariant manifold family tangent to the modal subspace and solved the computational-efficiency problem through the concept of spectral submanifolds (SSMs). This method \cite{(17)jain2022compute} can be combined with normal-form parametrisation to directly extract the forced vibration response curve. Furthermore, Cenedese and Haller \cite{(18)cenedese2020conservative} mathematically analysed the crucial relationship between the forced-damped backbone curves and the backbone curve of the corresponding conservative system. They derived a method for obtaining the former from the latter. An important corollary of their study---the phase lag quadrature criterion---theoretically validates the method used by Peeters et al. \cite{(19)peeters2011dynamic} for nonlinear system identification. The backbone curve computation method derived from this criterion is comparative in our subsequent analysis. 

A limitation of the SSM method is that it cannot be applied to nonsmooth nonlinear systems. For solving NNMs in systems with dry friction interfaces, Krack \cite{(15)krack2015nonlinear} provided a computationally feasible solution for dNNMs by extending the periodic-motion definition to dissipative systems. Building upon this framework, Sun \cite{(20)sun2021extended} predicted the forced resonance backbone curve by integrating Krack's dNNMs approach with the energy-balance method.

\subsection{Submanifolds and continuation algorithms}
The method developed in this study leverages the geometric properties of the forced response surface, where the resonance backbone curve is formed by its ridges. The parameter continuation techniques of Kernéve and Doedel \cite{(21)kernevez1987optimization} are a core element of search strategies for finding extrema along a constrained manifold. Dankowicz et al. \cite{(22)dankowicz2013recipes} developed a convenient software framework for parameter continuation methods called COCO. Building on these foundations, Mingwu Li \cite{(23)li2024fast} applied SSMs and successive continuation strategies with adjoints \cite{li2018staged} for constrained optimization and provided an efficient method for calculating the ridges and trenches on smooth forced response surfaces.

This study adopts the COCO framework using parameter continuation to solve for the extremal curves on the response surface of a nonsmooth system with dry friction interfaces, corresponding to the forced resonance backbone curve. The primary contributions of this study are twofold. First, it establishes a framework that integrates the HBM for nonsmooth systems with parameter continuation. Second, within this framework, the analytical Hessian Tensor of friction contact elements is derived from Petrov’s contact model to enable rapid and robust prediction of the forced resonance backbone curve.

The remainder of this paper is organised as follows: Section 2 introduces the theoretical background of the proposed method and two comparative methods for computing the backbone curve: the dNNMs approach defined by Sun and the phase resonance method. In Section 3, we establish the theoretical framework for calculating the HBM solution of a nonconservative system via parameter continuation. Section 4 introduces the friction contact model and derives a corresponding analytical Hessian Tensor. Integrating this Tensor with the framework described in Section 3 provides the final computational method for the forced resonance backbone curve. Section 5 presents the numerical validation of the proposed method. A cantilever-beam finite-element model was first used to demonstrate its effectiveness. Subsequently, a classical lumped-parameter blade--damper--blade model was employed for comparative analysis against the dNNMs and phase resonance methods. Section 6 concludes the paper with a summary of the research and discussion regarding the limitations of the method.

\section{Dynamic model of structures with friction contact interfaces} \label{sec:method}
\subsection{Dynamic equation of systems with friction} \label{ssec:model}
The dynamic equation of motion for an n-DOF structural system with friction nonlinearity can be expressed as
\begin{equation}
\boldsymbol{M} \ddot{\boldsymbol{q}}+\boldsymbol{C} \dot{\boldsymbol{q}}+\boldsymbol{K}\boldsymbol{q}+\boldsymbol{f_{n l}}=\boldsymbol{f_{e x}},
\end{equation}
where$\boldsymbol{M}$, $\boldsymbol{C}$, and $\boldsymbol{K}$ are the mass, damping, and stiffness matrices; $\boldsymbol{q}$ is the displacement vector; $\boldsymbol{f_{n l}}$ is the nonlinear contact force vector; and $\boldsymbol{f_{e x}}$ is the external excitation force vector.

By applying the Fourier--Galerkin projection, periodic quantities with an angular frequency of $\omega$ are expressed as a truncated Fourier series:
\begin{equation}
\begin{gathered}
\boldsymbol{q}(t)=\sum_{n=1}^{N_{\mathrm{h}}} \boldsymbol{Q}^{cn} \cos \mathrm{n} \omega t+\boldsymbol{Q}^{sn} \sin n \omega t \\
\boldsymbol{f}_{\boldsymbol{n} \boldsymbol{l}}(t)=\sum_{n=1}^{N_{\mathrm{h}}} {\boldsymbol{F_{nl}}}^{cn} \cos \mathrm{n} \omega t+{\boldsymbol{F_{nl}}}^{sn} \sin n \omega t \\
\boldsymbol{f}_{\boldsymbol{e x}}(t)==\sum_{n=1}^{N_{\mathrm{h}}} {\boldsymbol{F_{ex}}}^{cn} \cos \mathrm{n} \omega t+{\boldsymbol{F_{ex}}}^{sn} \sin n \omega t,
\end{gathered}
\end{equation}
where $\boldsymbol{Q}$, $\boldsymbol{Fnl}$, and $\boldsymbol{Fex}$ are the vectors of Fourier coefficients for the displacement, nonlinear contact force, and external excitation force, respectively.
\begin{equation}
\begin{gathered}
\boldsymbol{Q}=\left[\boldsymbol{Q}^{c 1}  ;  \boldsymbol{Q}^{s 1} ; \cdots ; \boldsymbol{Q}^{c N_h} ; \boldsymbol{Q}^{s N_h}\right] \\
{\boldsymbol{F_{nl}}}=\left[{\boldsymbol{F_{nl}}}^{c 1}  ;  {\boldsymbol{F_{nl}}}^{s 1} ; \cdots ; {\boldsymbol{F_{nl}}}^{c N_h} ; {\boldsymbol{F_{nl}}}^{s N_h}\right] \\
{\boldsymbol{F_{ex}}}=\left[{\boldsymbol{F_{ex}}}^{c 1}  ;  {\boldsymbol{F_{ex}}}^{s 1} ; \cdots ; {\boldsymbol{F_{ex}}}^{c N_h} ; {\boldsymbol{F_{ex}}}^{s N_h}\right]
\end{gathered}
\end{equation}

Equation (1) can be solved using the multiharmonic balance method to obtain the steady-state periodic response.
\begin{equation}
\boldsymbol{D}(\omega) \cdot \boldsymbol{Q}+{\boldsymbol{F_{nl}}}(\boldsymbol{Q}, \omega)-{\boldsymbol{F_{ex}}}=0
\end{equation}

Here, $\boldsymbol{D}(\omega)$ denotes the dynamic stiffness matrix of the underlying linear system. Because the contact force is independent of the frequency, the vector of nonlinear forces can be expressed in the simplified form ${\boldsymbol{F_{nl}}}(\boldsymbol{Q})$.
\begin{equation}
\resizebox{0.98\textwidth}{!}{
$\boldsymbol{D}(\omega)=\left[\begin{array}{ccccc}
\boldsymbol{K}-(\omega)^2 \boldsymbol{M} & \omega \boldsymbol{C} & \cdots & \mathbf{0} & \mathbf{0} \\
-\omega \boldsymbol{C} & \boldsymbol{K}-(\omega)^2 \boldsymbol{M} & \boldsymbol{M} & \cdots & \mathbf{0} \\
\cdots & \cdots & \cdots & \cdots & \mathbf{0} \\
0 & 0 & \cdots \boldsymbol{K}-\left(N_{\mathrm{h}} \omega\right)^2 \boldsymbol{M} & N_{\mathrm{h}} \omega \boldsymbol{C} & \\
0 & 0 & \cdots & -N_{\mathrm{h}} \omega \boldsymbol{C} & \boldsymbol{K}-\left(N_{\mathrm{h}} \omega\right)^2 \boldsymbol{M}
\end{array}\right]$
}
\end{equation}

Equation (4) represents the general form of the governing equation for calculating the forced vibration response of a nonlinear structure. 

A static/dynamic decoupled method is employed in this study, where the static component is excluded from the iterative process. The presented theory is also valid for coupled solution methods. For a detailed comparison of the two approaches, the reader is referred to \cite{(24)firrone2011effect}.
\subsection{Forced backbone curve computation via dNNMs} \label{ssec:Comparison method}
According to \cite{(15)krack2015nonlinear}, dNNMs replace external excitation by introducing an artificial negative modal damping factor $\xi$ to inject energy into a system. The damping factor is related to the modal damping ratio $2 \zeta \omega_0=\xi$. The resulting excitation force $\xi\boldsymbol{M}\dot{\boldsymbol{q}}$, which is proportional to the modal velocity, is inherently in phase with the velocity and thus automatically satisfies the phase resonance criterion. To characterise the energy-dependent properties of the NNM, the dNNMs introduce the modal amplitude $\alpha$ and define $psi$ as the corresponding modal eigenvector.
\begin{equation}
\boldsymbol{\psi}=\left[\boldsymbol{\psi}^{c 1} ; \boldsymbol{\psi}^{s 1} ; \ldots ; \boldsymbol{\psi}^{c n} ; \boldsymbol{\psi}^{s n} ; \ldots ; \boldsymbol{\psi}^{c N_{\mathrm{h}}} ; \boldsymbol{\psi}^{s N_{\mathrm{h}}}\right]
\end{equation}
Here, $\boldsymbol{\psi^{c n}}$ and $\boldsymbol{\psi^{s n}}$ are the sine and cosine coefficient vectors of the $n^{th}$ harmonic of the modal eigenvector, respectively. The modal displacement of the system is given as $\boldsymbol{Q}=\alpha \cdot \boldsymbol{\psi}$. Substituting the displacement vector in Equation (5) with the modal displacement yields the following equation:
\begin{equation}
\boldsymbol{\bar{D}}\left(\omega_0\right) \cdot \alpha \cdot \boldsymbol{\psi}+\boldsymbol{F_{nl}}(\alpha \cdot \boldsymbol{\psi})=0,
\end{equation}
where $\omega_0$ represents the modal frequency and $\boldsymbol{\bar{D}}$ is the dynamic stiffness matrix incorporating negative modal damping.
\begin{equation}
\resizebox{0.98\textwidth}{!}{
$\boldsymbol{D}\left(\omega_0\right)=\left[\begin{array}{ccccc}
\boldsymbol{K}-\left(\omega_0\right)^2 \boldsymbol{M} & \omega_0 \boldsymbol{C}+\xi \omega_0 \mathbf{M} & \cdots & \mathbf{0} & \mathbf{0} \\
-\omega_0 \boldsymbol{C}-\xi \omega_0 \mathbf{M} & \boldsymbol{K}-\left(\omega_0\right)^2 \boldsymbol{M} & \boldsymbol{M} & \cdots & \mathbf{0} \\
\cdots & \cdots & \cdots & \cdots & \mathbf{0} \\
0 & 0 & \cdots \boldsymbol{K}-\left(N_{\mathrm{h}} \omega_0\right)^2 \boldsymbol{M} & N_{\mathrm{h}} \omega_0 \boldsymbol{C}+\xi N_{\mathrm{h}} \omega_0 \mathbf{M} & \\
0 & 0 & \cdots & -N_{\mathrm{h}} \omega_0 \boldsymbol{C} & \boldsymbol{K}-\left(N_{\mathrm{h}} \omega_0\right)^2 \boldsymbol{M}
\end{array}\right]$
}
\end{equation}
As the modal damping ratio and frequency are unknown variables to be solved, two additional constraints---a fundamental harmonic mass normalisation and a phase normalisation condition---must be imposed to ensure a unique solution to the equations.
\begin{equation}
\boldsymbol{\psi}^{c 1^{\mathrm{T}}} \boldsymbol{M} \boldsymbol{\psi}^{c 1}+\boldsymbol{\psi}^{ s 1^{\mathrm{T}}} \boldsymbol{M} \boldsymbol{\psi}^{s 1}=1, \psi^{s 1}{ }_k=0
\end{equation}
Here, the superscript T denotes the transpose operation, and $\psi^{s 1}{ }_k$ represents the cosine component of the fundamental harmonic for the $k^{th}$ DOF.

Following the approach of \cite{(20)sun2021extended}, the resonance peak of the forced vibration response is assumed to be equal to the nonlinear modal displacement described above. Accordingly, by considering only the fundamental harmonic, an energy-balance criterion for the NNM can be established to determine the equivalent external excitation.

\begin{equation}
E_{\mathrm{d}}=E_{\mathrm{f}}
\end{equation}
Here, $E_{\mathrm{d}}$ represents the energy injected or dissipated by the modal damping term, and $E_{\mathrm{f}}$ represents the energy supplied by the equivalent external excitation.

\begin{equation}
\begin{aligned}
& E_{\mathrm{d}}=
\int_0^{2 \pi / \omega_0} 2\zeta\omega_0 {\dot{\boldsymbol{q}}}^{T} \boldsymbol{M} \dot{\boldsymbol{q}}d t \\
& E_{\mathrm{f}}= \int_0^{2 \pi / \omega_0} \boldsymbol{f_{ex}}^T\dot{\boldsymbol{q}} d t
\end{aligned}
\end{equation}

After integration, we obtain
\begin{equation}
\begin{aligned}
& E_{\mathrm{d}}=\sum_{n=1}^{N_{\mathrm{h}}}{ -2\pi\zeta\left(n \omega_0\right)^2 \left(\|\mathbf{M} {\boldsymbol{Q}^{c n}}^{\odot 2}\|_1 + \|\mathbf{M} {\boldsymbol{Q}^{ s n}}^{\odot 2}\|_1\right) }  \\
& E_{\mathrm{f}}=\sum_{n=1}^{N_{\mathrm{h}}}{ \pi\left(n \omega_0\right) \left(\|{\boldsymbol{F_{ex}}}^{s n}{\boldsymbol{Q}^{c n}}\|_1 + \|{\boldsymbol{F_{ex}}}^{c n}{\boldsymbol{Q}^{s n}}\|_1\right) },
\end{aligned}
\end{equation}
where$\odot 2$ denotes the element-wise square of the vector. ${\boldsymbol{F_{ex}}{}^{s n}}$ and ${\boldsymbol{F_{ex}}{}^{c n}}$ are the sine and cosine coefficient vectors of the $n^{th}$ harmonic of ${\boldsymbol{F_{ex}}}$, respectively.

Solving Equation (10) yields the equivalent excitation force vector, which enables the construction of the resonance backbone curve at different excitation levels.
\subsection{Forced backbone curve computation via phase resonance method} 
For linear systems, a 90° phase lag between excitation and response is a well-established criterion for identifying resonance (also termed phase resonance). This phase-lag quadrature criterion was extended by Peeters et al. \cite{(19)peeters2011dynamic} to monophase NNM motions of nonlinear structures. Subsequently, George Haller \cite{(18)cenedese2020conservative} provided mathematical conditions under which approximate numerical and experimental approaches, such as energy balance and force appropriation, are justifiable. The phase-lag quadrature criterion was also adapted as a comparative approach to compute the forced resonance backbone curve for systems with friction contact interfaces. First, Equation (4) is rewritten as follows:
\begin{equation}
\boldsymbol{D}(\omega) \cdot \boldsymbol{Q}+\boldsymbol{F_{nl}}(\boldsymbol{Q})+\alpha \boldsymbol{F_{ex}}=0,
\end{equation}
where $\alpha$ is a scaling parameter representing the level of excitation force. For a single-point monoharmonic excitation, the external force vector in Equation (4) takes the following form:
\begin{equation}
\begin{gathered}
{\boldsymbol{F_{ex}}}=\left[{{\boldsymbol{F_{ex}}} ^{c 1}}^T , 0, \ldots, 0,\right]^T \\
{\boldsymbol{F_{ex}}}^{c 1}=[0, \ldots, 0, \overbrace{1}^{\text {kth }}, 0 \ldots, 0]^T.
\end{gathered}
\end{equation}
In other words, a sinusoidal excitation is applied to the k-th DOF of the system, and the phase-lag quadrature criterion is then imposed.
\begin{equation}
Q^{c 1}{ }_k=0
\end{equation}

Computing the forced response for a system with friction contact interfaces involves an AFT scheme and a continuation process. The detailed procedure is illustrated in Figure 1.
\begin{figure}
  \centering
  \includegraphics[width=0.8\textwidth]{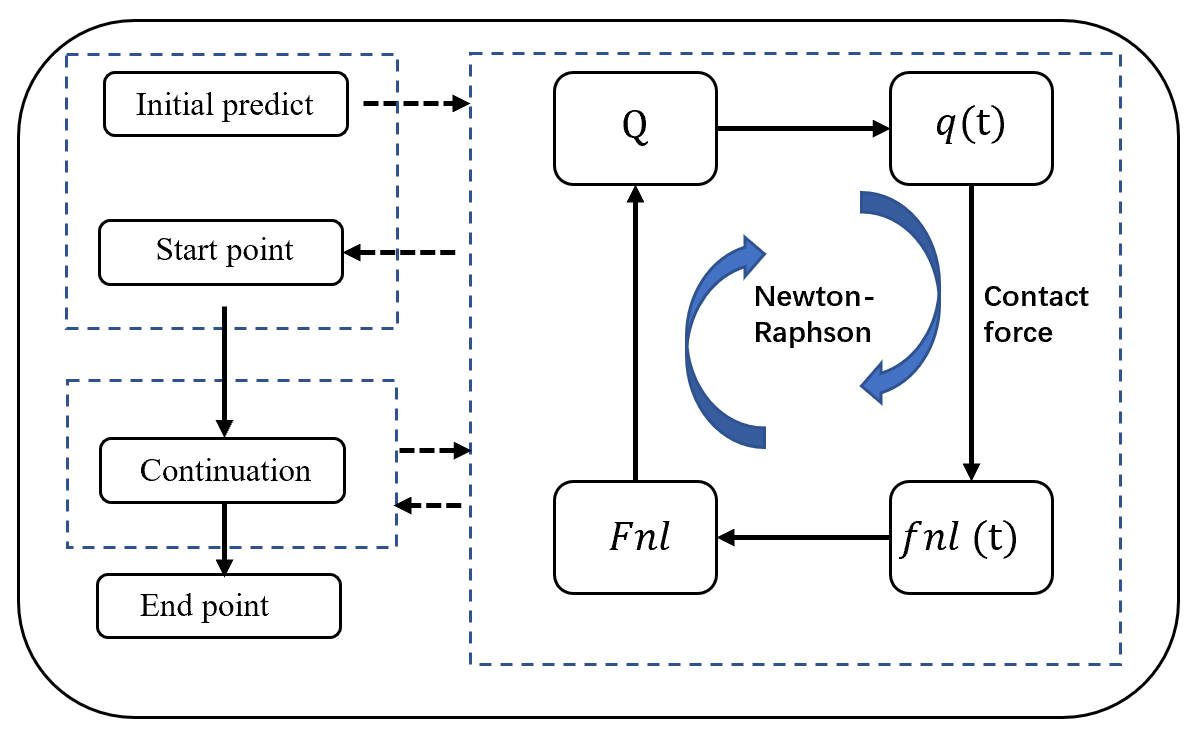}
  \caption{Schematic of the AFT method for forced vibration response computation}
\end{figure}

For dNNMs and the phase resonance method, a single iterative process can yield only one resonance backbone curve. Therefore, selecting the initial starting point is crucial for obtaining the backbone curve corresponding to the target nonlinear mode. When the excitation level is low, the friction interface is in a fully stuck state, under which the system degenerates into a linear state. Consequently, the resonant response of the corresponding linear mode in this stuck state is typically selected as the initial point for iteration. The range of the continuation parameter is then set as
\begin{equation}
\alpha \in\left[\alpha_{\min }, \alpha_{\max }\right].
\end{equation}
\section{Forced backbone curve tracing via parameter continuation} \label{sec:method}
\subsection{Response surface equation for forced vibration} \label{ssec:equation}
Parameter continuation techniques provide a robust framework for the global exploration and tracing of smooth solution manifolds in underdetermined systems of equations. A smooth manifold with multiple variables and parameters can be defined using the following equation:
\begin{equation}
\begin{aligned}
& \Phi(\boldsymbol{u}, \boldsymbol{\mu})=0 \\
& \Psi(\boldsymbol{\mu})-\boldsymbol{\mu_0}=0,
\end{aligned}
\end{equation}
where $\boldsymbol{u}$ denotes a collection of variables, $\boldsymbol{\mu}$ denotes a collection of parameters, and $\Phi(u, \mu)$. Here, $\Phi(\boldsymbol{u}, \boldsymbol{\mu})=0$ can be regarded as the equality constraint of the manifold, and $\Psi(\boldsymbol{\mu})-\boldsymbol{\mu_0}=0$ can be regarded as the constraint on the corresponding parameters. Correspondingly, Equation (4) can be rewritten as follows:
\begin{equation}
\begin{gathered}
R(\boldsymbol{Q}, \omega, \alpha) \\
\binom{\alpha-\alpha_0}{\omega-\omega_0}=0,
\end{gathered}
\end{equation}
where
\begin{equation}
\begin{aligned}
R(\boldsymbol{Q}, \omega, \alpha) & =D(\omega) \cdot \boldsymbol{Q}+\boldsymbol{F_{n l}}(\boldsymbol{Q})-\alpha \boldsymbol{F_{e x}} \\
& \alpha_0 \in\left[\alpha_{\min }, \alpha_{\max }\right] \\
& \omega_0 \in\left[\omega_{\min }, \omega_{\max }\right].
\end{aligned}
\end{equation}

The zero locus of equation (19) defines the forced response surface of the nonlinear system as a function of vibration frequency and excitation amplitude. Geometrically, the ridges of the response surface, that is, the trajectories of its local extrema, are equivalent to forced resonance backbone curves. Thus, the problem of solving the backbone curves is transformed into the problem of finding the ridges of the response surface. For example, in a 2-DOF oscillator with cubic nonlinearity \cite{(25)krack2019harmonic}, as illustrated in Figure 2, the two ridges formed by connecting the local extrema on the response surface correspond to the backbone curves of the system's two nonlinear modes. This correspondence between backbone curves and ridges remains valid for nonsmooth nonlinearities like friction. Subsequently, the specific method for computing the backbone curves for systems with friction contact interfaces is introduced.
\begin{figure}
  \centering
  \includegraphics[width=0.8\textwidth]{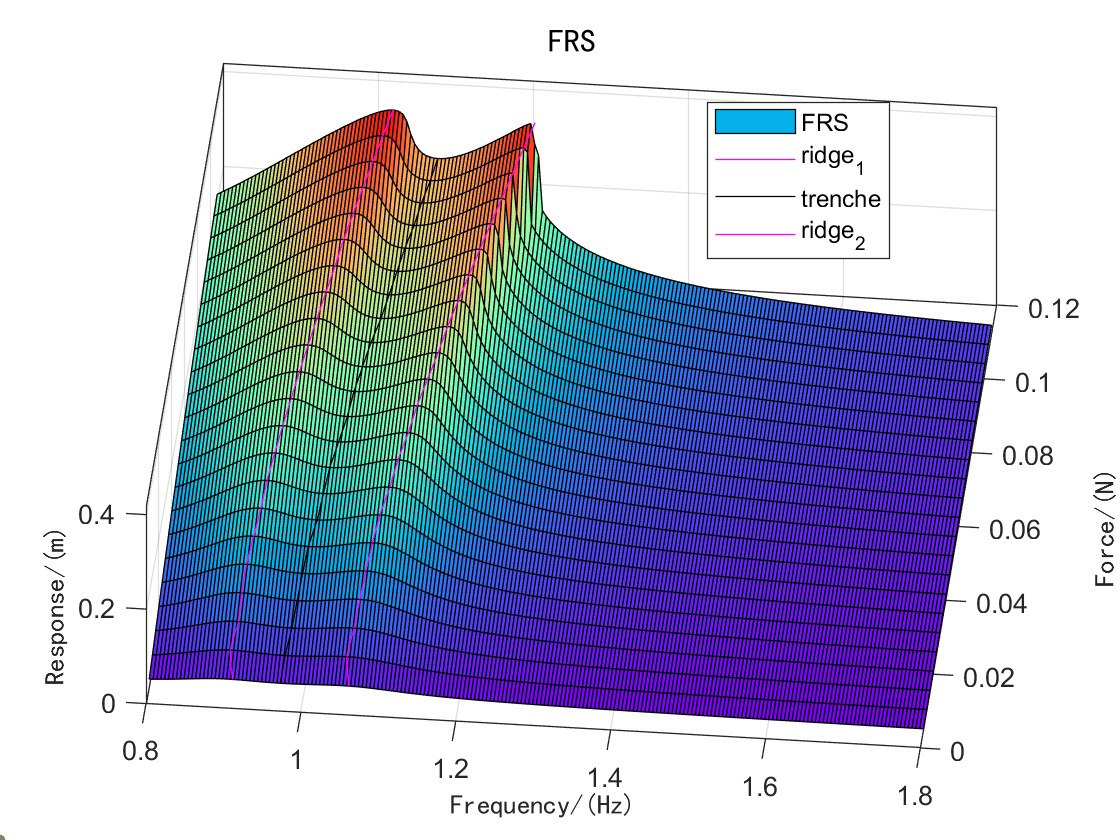}
  \caption{Ridges and trenches on the forced response surface}
\end{figure}
\subsection{Backbone curve tracing via parameter continuation method} \label{ssec:method}
Parameter continuation techniques combined with Lagrange multipliers can be effectively deployed to trace extreme values along a constraint manifold.

First, we define an amplitude-monitoring function. According to Parseval's theorem, the magnitude of the vibration energy for the k-th DOF can be defined using the following expression:
\begin{equation}
\begin{aligned}
A m=\frac{\sqrt{2}}{2} \sqrt{\sum_{n=1}^{N_{\mathrm{h}}} {Q}^{c n}{ }_k{ }^2+{Q}^{s n}{ }_k{ }^2}=\frac{\sqrt{2}}{2}\|\boldsymbol{Q_k}\| \\
\boldsymbol{Q_k}=\left[{{Q}^{c 1}{ }_k} ,{{Q}^{s 1}{ }_k}  \ldots, {{Q}^{c n}{ }_k} ,{{Q}^{s n}{ }_k},\ldots, {{Q}^{c N_h}{ }_k} ,{{Q}^{s N_h}{ }_k}\right]^T.
\end{aligned}
\end{equation}

By introducing auxiliary parameters, the Lagrangian for Equations (18) and (20) can be written in the following form:
\begin{equation}
\begin{aligned}
L= & A m+\eta_{A m}\left(A m-A m_0\right)+ \\
& \eta_\alpha\left(\alpha-\alpha_0\right)+\eta_\omega\left(\omega-\mu_\omega\right)+\lambda^{\top} \boldsymbol{h},
\end{aligned}
\end{equation}
where $\eta_{A m}, \eta_\alpha, \eta_\omega, \lambda$ are the corresponding Lagrange multipliers. To find the extremal points, the partial derivatives of the Lagrangian function with respect to the variables $\boldsymbol{Q}, \omega, \alpha$ and the Lagrange multipliers are set to zero, which yields
\begin{equation}
R e=\left\{\begin{array}{l}
A m-A m_0=0, \alpha-\alpha_0=0, \omega-\omega_0=0, R=\mathbf{0}, \\
\frac{\partial A m}{\partial \boldsymbol{Q}} \eta_{A m}+\left(\frac{\partial \boldsymbol{R}}{\partial \boldsymbol{Q}}\right)^{\top} \lambda=\mathbf{0}, \\
\frac{\partial A m}{\partial \omega} \eta_{A m}+\eta_\omega+\left(\frac{\partial \boldsymbol{R}}{\partial \omega}\right)^{\top} \lambda=0, \\
\frac{\partial A m}{\partial \alpha} \eta_{A m}+\eta_\alpha+\left(\frac{\partial \boldsymbol{R}}{\partial \alpha}\right)^{\top} \lambda=0,
\end{array}\right.
\end{equation}
where
\begin{equation}
\begin{gathered}
\frac{\partial A m}{\partial \boldsymbol{Q_k}}=\frac{\sqrt{2}}{2} \frac{\boldsymbol{Q_k}}{\|\boldsymbol{Q_k}\|} \\
\left(\frac{\partial \boldsymbol{R}}{\partial \boldsymbol{Q}}\right)=\boldsymbol{D}(\omega)+\frac{\partial \operatorname{\boldsymbol{F_{nl}}}(\boldsymbol{Q})}{\partial \boldsymbol{Q}} \\
\left(\frac{\partial \boldsymbol{R}}{\partial \omega}\right)=\frac{\partial \boldsymbol{D}(\omega)}{\partial \omega} \\
\left(\frac{\partial \boldsymbol{R}}{\partial \alpha}\right)=\boldsymbol{F_{ex}}.
\end{gathered}
\end{equation}

According to the approach proposed by Li \cite{(23)li2024fast}, we outline the parameter continuation process as follows:
\begin{enumerate}[1)]
\item First, for a given excitation parameter $\alpha_{\min }$, the frequency $\omega_{0 }$ is selected as the continuation parameter. Continuation is performed over an interval that includes the nonlinear modal frequency. The forced response curve (FRC) is computed within this range. All Lagrange multipliers are set to zero at this stage, which automatically satisfies the adjoint equations. The COCO framework is then used to automatically locate the fold points on the FRC, which correspond to the maximum amplitudes.
\item Subsequently, a second continuation is performed with $\eta_{A m}$ as the continuation parameter, starting from the previously detected fold point. The parameter is varied until $\eta_{A m}=1$ to trace the secondary branch. During this stage, only the Lagrange multipliers $\eta_{A m}, \eta_\alpha, \eta_\omega, \lambda$ vary linearly, whereas the design variables $\boldsymbol{Q}, \omega ,\alpha $ are held constant.
\item Subsequently, using $\alpha$ as the continuation parameter, the forced vibration backbone curve is traced over the interval $\alpha_0 \in\left[\alpha_{\min }, \alpha_{\max }\right]$.
\item As demonstrated in the numerical analysis in Section 6, Step (3) of the procedure may encounter numerical singularities. When the continuation process detects a numerical singularity point $\alpha_{mx}$, an event is triggered at this location within the COCO framework to initialise a new continuation problem for $\alpha_0 \in\left[\alpha_{\operatorname{mx}}+\epsilon, \alpha_{\max }\right]$. The procedure then resumes by reactivating Steps (1)–-(3) at a location slightly downstream of this point, continuing until $\alpha=\alpha_{\max }$.
\end{enumerate}
\section{Analytical Hessian Tensor of contact elements } \label{sec:method}
\subsection{Friction contact elements} \label{ssec:model}
Solving Equations (7), (13), and (22) requires the modelling of the friction contact elements. In this study, Yang's friction model \cite{(3)yang1998stick} was adopted, and the method \cite{(9)petrov2003analytical} was used to compute the frequency-domain nonlinear contact forces and stiffness matrix for these contact elements. In this subsection, we review previous research to provide the necessary theoretical basis for deriving the Hessian Tensor of contact elements in the subsequent section.

A schematic of the contact model for a single-node pair is shown in Figure 3. The relative tangential and normal motions at the contact interface are denoted as $x$ and $y$, respectively. The parameters for the contact element include the coefficient of friction $\mu$, tangential contact stiffness $k_x$, normal contact stiffness $k_y$, and normal contact load $N_0$.
\begin{figure}
  \centering
  \includegraphics[width=0.8\textwidth]{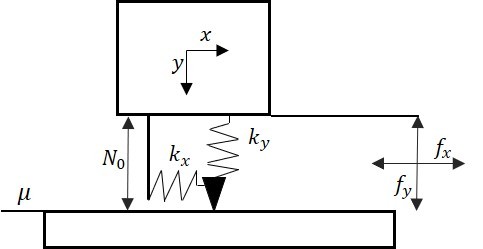}
  \caption{Contact model}
\end{figure}

Accounting for the variation in the normal force due to normal motion, the expressions for the nonlinear tangential contact forces for all possible contact states can be formulated as follows:
\begin{equation}
f_x=\left\{\begin{array}{cc}
f_x^0+k_x\left(x-x_0\right) & \text { for stick } \\
\xi \mu f_y & \text { for slip } \\
0 & \text { for separation }
\end{array}\right.
\end{equation}

The normal contact forces for all possible contact states can be written as
\begin{equation}
f_y=\left\{\begin{array}{cc}
N_0+k_y y & \text { for contact } \\
0 & \text { for separation }
\end{array}\right.
\end{equation}

where $\xi$ represents the sign of the tangential force evaluated at the instant slip begins; $x_0$ and ${f}_{{x}}{ }^0 = \xi \mu f_y\left(\tau_{\text {stick }}\right)$ represent the initial tangential displacement and force at the onset of the stick state, respectively.

Computing the frequency-domain contact forces and the stiffness matrix of contact elements requires the determination of the time-of-contact state transitions (e.g., from stick to slip or from slip to separation). The process starts at an instant of sticking, at which time the tangential displacement is $x_0$ and the tangential force is ${f}_{{x}}{ }^0$. The subsequent transition times are determined using the conditions outlined below.

The criterion for the transition from stick to slip is as follows:
\begin{equation}
f_x^0+k_x\left(x(\tau)-x_0\right)=\xi \mu\left(N_0+k_y y(\tau)\right).
\end{equation}

The criterion for the transition from slip to stick is as follows:
\begin{equation}
\begin{aligned}
& \xi k_x \dot{x}(\tau)=\mu k_y \dot{y}(\tau) \\
& \xi k_x \ddot{x}(\tau)<\mu k_y \ddot{y}(\tau).
\end{aligned}
\end{equation}

The criterion for the transition from contact to separation is
\begin{equation}
\begin{aligned}
& N_0+k_y y(\tau)=0 \\
& \dot{y}(\tau)>0.
\end{aligned}
\end{equation}

The criterion for the transition from separation to contact is
\begin{equation}
\begin{aligned}
& N_0+k_y y(\tau)=0 \\
& \dot{y}(\tau)<0.
\end{aligned}
\end{equation}

The reader is referred to \cite{(9)petrov2003analytical}. 

The Fourier coefficients of the contact force can be obtained by integrating over one period of vibration, as follows:
\begin{equation}
\left\{\begin{array}{l}
\boldsymbol{F}_x \\
\boldsymbol{F}_y
\end{array}\right\}=\frac{1}{\pi} \sum_{j=1}^{n_\tau} \int_{\tau_j}^{\tau_{j+1}}\left\{\begin{array}{l}
\boldsymbol{H}_{+}(\tau) f_x \\
\boldsymbol{H}_{+}(\tau) f_y
\end{array}\right\} d \tau=\sum_{j=1}^{n_\tau}\left\{\begin{array}{l}
\boldsymbol{J}_x^{(j)} \\
\boldsymbol{J}_y^{(j)}
\end{array}\right\},
\end{equation}
where
\begin{equation}
\boldsymbol{H}_{+}=\left\{\cos \tau, \sin \tau, \ldots, \cos N_{\mathrm{h}} \tau, \sin N_{\mathrm{h}} \tau\right\}^T
\end{equation}
\begin{equation}
\boldsymbol{J}_x^{(j)}= \begin{cases}k_x \boldsymbol{W}_j \boldsymbol{X}+c_j \boldsymbol{w}_j & \text { stick } \\ \xi_\mu\left(N_0 \boldsymbol{w}_j+k_y \boldsymbol{W}_j \boldsymbol{Y}\right) & \text { slip } \\ \mathbf{0} & \text { separation }\end{cases}
\end{equation}
\begin{equation}
\boldsymbol{J}_y^{(j)}= \begin{cases}N_0 \boldsymbol{w}_j+k_y \boldsymbol{W}_j \boldsymbol{Y} & \text { contact } \\ \mathbf{0} & \text { separation }\end{cases}
\end{equation}
\begin{equation}
\underset{\left(2 N_h \times 2 N_h\right)}{\boldsymbol{W}_j}=\frac{1}{\pi} \int_{\tau_j}^{\tau_{j+1}} \boldsymbol{H}_{+}(\tau) \boldsymbol{H}_{-}^T(\tau) d \tau ; \underset{\left(2 N_h \times 1\right)}{\boldsymbol{w}_j}=\frac{1}{\pi} \int_{\tau_j}^{\tau_{j+1}} \boldsymbol{H}_{+}(\tau) d \tau
\end{equation}
\begin{equation}
c_j=f_x^0\left(\tau_j\right)-k_x x\left(\tau_j\right)
\end{equation}
\begin{equation}
\boldsymbol{H}_{-}=\left\{\cos \tau, \sin \tau, \ldots, \cos N_{\mathrm{h}} \tau, \sin N_{\mathrm{h}} \tau\right\}^T.
\end{equation}
In contrast to the original method employed in \cite{(9)petrov2003analytical}, the static component was not considered in this study.

The stiffness matrix for the contact element is expressed as follows:
\begin{equation}
\boldsymbol{K}_f=\left[\begin{array}{cc}
\frac{\partial \boldsymbol{F}_x}{\partial \boldsymbol{X}} & \frac{\partial \boldsymbol{F}_x}{\partial \boldsymbol{Y}} \\
\mathbf{0} & \frac{\partial \boldsymbol{F}_y}{\partial \boldsymbol{Y}}
\end{array}\right]=\sum_{j=1}^{n_\tau}\left[\begin{array}{cc}
\frac{\partial \boldsymbol{J}_x^{(j)}}{\partial \boldsymbol{X}} & \frac{\partial \boldsymbol{J}_x^{(j)}}{\partial \boldsymbol{Y}} \\
\mathbf{0} & \frac{\partial \boldsymbol{J}_y^{(j)}}{\partial \boldsymbol{Y}}
\end{array}\right],
\end{equation}
where
\begin{equation}
\frac{\partial \boldsymbol{J}_x^{(j)}}{\partial \boldsymbol{X}}= \begin{cases}k_x \boldsymbol{W}_j+\boldsymbol{w}_j\left(\frac{\partial c_j}{\partial \boldsymbol{X}}\right)^T & \text { stick } \\ \mathbf{0} & \text { slip } \\ \mathbf{0} & \text { separation }\end{cases}
\end{equation}
\begin{equation}
\frac{\partial \boldsymbol{J}_x^{(j)}}{\partial \boldsymbol{Y}}= \begin{cases}\boldsymbol{w}_j\left(\frac{\partial c_j}{\partial \boldsymbol{Y}}\right)^T & \text { stick } \\ \xi \mu k_y \boldsymbol{W}_j & \text { slip } \\ \mathbf{0} & \text { separation }\end{cases}
\end{equation}
\begin{equation}
\frac{\partial \boldsymbol{J}_y^{(j)}}{\partial \boldsymbol{Y}}= \begin{cases}k_y \boldsymbol{W}_j & \text { contact } \\ \mathbf{0} & \text { separation }\end{cases}
\end{equation}
\begin{equation}
\begin{aligned}
& \frac{\partial c_j}{\partial \boldsymbol{X}}=\xi \mu k_y \dot{y}\left(\tau_j\right) \frac{\partial \tau_j}{\partial \boldsymbol{X}}-k_x\left(\boldsymbol{H}_{-}\left(\tau_j\right)+\dot{x}\left(\tau_j\right) \frac{\partial \tau_j}{\partial \boldsymbol{X}}\right) \\
& \frac{\partial c_j}{\partial \boldsymbol{Y}}=\xi \mu k_y\left(\boldsymbol{H}_{-}\left(\tau_j\right)+\dot{y}\left(\tau_j\right) \frac{\partial \tau_j}{\partial \boldsymbol{Y}}\right)-k_x \dot{x}\left(\tau_j\right) \frac{\partial \tau_j}{\partial \boldsymbol{Y}}.
\end{aligned}
\end{equation}
Here, $\frac{\partial \tau_j}{\partial \boldsymbol{X}}$ and $\frac{\partial \tau_j}{\partial \boldsymbol{Y}}$ represent the partial derivatives of the sticking time $\tau_j$ with respect to the tangential $x$ and normal $y$ displacements, respectively. Their values are determined by considering two separate cases.
When $\tau_j$ is the transition time from the slip state to the stick state,
\begin{equation}
\begin{aligned}
& \frac{\partial \tau_{\text {stick }}}{\partial \boldsymbol{X}}=\frac{-\xi k_x}{\xi k_x \ddot{x}\left(\tau_{\text {stick }}\right)-\mu k_y \ddot{y}\left(\tau_{\text {stick }}\right)} \dot{\boldsymbol{H}}_{-}\left(\tau_{\text {stick }}\right) \\
& \frac{\partial \tau_{\text {stick }}}{\partial \boldsymbol{Y}}=\frac{\mu k_y}{\xi k_x \ddot{x}\left(\tau_{\text {stick }}\right)-\mu k_y \ddot{y}\left(\tau_{\text {stick }}\right)} \dot{\boldsymbol{H}}_{-}\left(\tau_{\text {stick }}\right).
\end{aligned}
\end{equation}

When $\tau_j$ is the transition time from the separation state to the stick state,
\begin{equation}
\begin{gathered}
\frac{\partial \tau_{\text {stick }}}{\partial \boldsymbol{X}}=\mathbf{0} \\
\frac{\partial \tau_{\text {stick }}}{\partial \boldsymbol{Y}}=-\frac{1}{\dot{y}\left(\tau_{\text {stick }}\right)} \boldsymbol{H}_{-}\left(\tau_{\text {stick }}\right).
\end{gathered}
\end{equation}
\subsection{Hessian Tensor of contact elements} \label{ssec:model}
Because Equation (22) contains the first derivative of the nonlinear force with respect to the displacement Fourier coefficients, the computation of its Jacobian matrix requires the stiffness matrix of the contact element and its analytical Hessian Tensor. Here, the ‘Hessian’ is a collection of $4 N_h $ matrices. This collection can be understood as the components of a third-order tensor representing the partial derivative of the stiffness matrix (the force Jacobian) of the contact elements with respect to the displacement vector $\left[\boldsymbol{X};\boldsymbol{Y}\right]$. Each Hessian matrix corresponds to the derivative of one row of the stiffness matrix.

The expression for the i-th Hessian matrix, which is obtained by differentiating the i-th row $\boldsymbol{K}_f(i,:)$ of the Jacobian matrix $\boldsymbol{K}_f$, is given as
\begin{equation}
\operatorname{Hes}_i=\left[\frac{\partial \boldsymbol{K}_f(i,:)^T}{\partial \boldsymbol{X}}, \frac{\partial \boldsymbol{K}_f(i,:)^T}{\partial \boldsymbol{Y}}\right]=\sum_{j=1}^{n_\tau}\left[\begin{array}{ll}
\frac{\partial \frac{\partial \boldsymbol{J}_x^{(j)}}{\partial \boldsymbol{X}}(i,:)^T}{\partial \boldsymbol{X}} & \frac{\partial \frac{\partial \boldsymbol{J}_x^{(j)}}{\partial \boldsymbol{X}}(i,:)^T}{\partial \boldsymbol{Y}} \\
\frac{\partial \frac{\partial \boldsymbol{J}_x^{(j)}}{\partial \boldsymbol{Y}}(i,:)^T}{\partial \boldsymbol{X}} & \frac{\partial \frac{\partial \boldsymbol{J}_x^{(j)}}{\partial \boldsymbol{Y}}(i,:)^T}{\partial \boldsymbol{Y}}
\end{array}\right],
\end{equation}
where the second-order partial derivative matrix within the larger matrix is obtained by differentiating Equations (37) and (38).
\begin{equation}
\begin{aligned}
& \frac{\partial \frac{\partial \boldsymbol{J}_x^{(j)}}{\partial \boldsymbol{X}}(i,:)^T}{\partial \boldsymbol{X}} \\
& = \begin{cases}k_x\left(\frac{\partial \boldsymbol{W}_j(i,:)^T}{\partial \tau_{j+1}} \frac{\partial \tau_{j+1}}{\partial \boldsymbol{X}}-\frac{\partial \boldsymbol{W}_j(i,:)^T}{\partial \tau_j} \frac{\partial \tau_j}{\partial \boldsymbol{X}}\right)+\boldsymbol{w}_j(i) \frac{\partial^2 c_j}{\partial \boldsymbol{X}^2}+ \\
\left(\frac{\partial \boldsymbol{w}_j(i)}{\partial \tau_{j+1}} \frac{\partial \tau_{j+1}}{\partial \boldsymbol{X}}-\frac{\partial \boldsymbol{w}_j(i)}{\partial \tau_j} \frac{\partial \tau_j}{\partial \boldsymbol{X}}\right) \frac{\partial c_j^T}{\partial \boldsymbol{X}} & \text { stick } \\
\mathbf{0} & \text { slip } \\
\mathbf{0} & \text { separation }\end{cases}
\end{aligned}
\end{equation}
\begin{equation}
\begin{aligned}
& \frac{\partial \frac{\partial \boldsymbol{J}_x^{(j)}}{\partial \boldsymbol{X}}(i,:)^T}{\partial \boldsymbol{Y}} \\
& = \begin{cases}k_x\left(\frac{\partial \boldsymbol{W}_j(i,:)^T}{\partial \tau_{j+1}} \frac{\partial \tau_{j+1}}{\partial \boldsymbol{Y}}-\frac{\partial \boldsymbol{W}_j(i,:)^T}{\partial \tau_j} \frac{\partial \tau_j}{\partial \boldsymbol{Y}}\right)+\boldsymbol{w}_j(i) \frac{\partial^2 c_j}{\partial \boldsymbol{X} \partial \boldsymbol{Y}}+\\
\left(\frac{\partial \boldsymbol{w}_j(i)}{\partial \tau_{j+1}} \frac{\partial \tau_{j+1}}{\partial \boldsymbol{Y}}-\frac{\partial \boldsymbol{w}_j(i)}{\partial \tau_j} \frac{\partial \tau_j}{\partial \boldsymbol{Y}}\right) \frac{\partial c_j^T}{\partial \boldsymbol{X}} & \text { stick } \\
\mathbf{0} & \text { slip } \\
\mathbf{0} & \text { separation }\end{cases}
\end{aligned}
\end{equation}
\begin{equation}
\frac{\partial \frac{\partial \boldsymbol{J}_x^{(j)}}{\partial \boldsymbol{Y}}(i,:)^T}{\partial \boldsymbol{X}}= \begin{cases}\boldsymbol{w}_j(i) \frac{\partial^2 c_j}{\partial \boldsymbol{Y} \partial \boldsymbol{X}}+\left(\frac{\partial \boldsymbol{w}_j(i)}{\partial \tau_{j+1}} \frac{\partial \tau_{j+1}}{\partial \boldsymbol{X}}-\frac{\partial \boldsymbol{w}_j(i)}{\partial \tau_j} \frac{\partial \tau_j}{\partial \boldsymbol{X}}\right) \frac{\partial c_j^T}{\partial \boldsymbol{Y}} & \text { stick } \\ \xi \mu k_y\left(\frac{\partial \boldsymbol{W}_j(i,:)^T}{\partial \tau_{j+1}} \frac{\partial \tau_{j+1}}{\partial \boldsymbol{X}}-\frac{\partial \boldsymbol{W}_j(i,:)^T}{\partial \tau_j} \frac{\partial \tau_j}{\partial \boldsymbol{X}}\right) & \text { slip } \\ \mathbf{0} & \text { separation }\end{cases}
\end{equation}
\begin{equation}
\frac{\partial \frac{\partial \boldsymbol{J}_x^{(j)}}{\partial \boldsymbol{Y}}(i,:)^T}{\partial \boldsymbol{Y}}= \begin{cases}\boldsymbol{w}_j(i) \frac{\partial^2 c_j}{\partial \boldsymbol{Y} \partial \boldsymbol{Y}}+\left(\frac{\partial \boldsymbol{w}_j(i)}{\partial \tau_{j+1}} \frac{\partial \tau_{j+1}}{\partial \boldsymbol{Y}}-\frac{\partial \boldsymbol{w}_j(i)}{\partial \tau_j} \frac{\partial \tau_j}{\partial \boldsymbol{Y}}\right) \frac{\partial c_j^T}{\partial \boldsymbol{Y}} & \text { stick } \\ \xi \mu k_y\left(\frac{\partial \boldsymbol{W}_j(i,:)^T}{\partial \tau_{j+1}} \frac{\partial \tau_{j+1}}{\partial \boldsymbol{Y}}-\frac{\partial \boldsymbol{W}_j(i,:)^T}{\partial \tau_j} \frac{\partial \tau_j}{\partial \boldsymbol{Y}}\right) & \text { slip } \\ \mathbf{0} & \text { separation }\end{cases}
\end{equation}

Here, $\frac{\partial \boldsymbol{W}_j(i,:)^T}{\partial \tau}$ is the derivative of $\boldsymbol{{W}_j(i,:)}$ with respect to the limits of integration:
\begin{equation}
\begin{gathered}
\frac{\partial \boldsymbol{W}_j}{\partial \tau}=\frac{1}{\pi}\left(\boldsymbol{H}_{+}(\tau) \boldsymbol{H}_{-}^T(\tau)\right) ; \\
\frac{\partial \boldsymbol{w}_j}{\partial \tau}=\frac{1}{\pi} \boldsymbol{H}_{+}(\tau).
\end{gathered}
\end{equation}
$\frac{\partial^2 c_j}{\partial \boldsymbol{X}^2}$, $\frac{\partial^2 c_j}{\partial \boldsymbol{X} \partial \boldsymbol{Y}}$, $\frac{\partial^2 c_j}{\partial \boldsymbol{X} \partial \boldsymbol{Y}}$, and $\frac{\partial^2 c_j}{\partial \boldsymbol{Y}^2}$ are obtained by taking the partial derivatives of Equation (40) with respect to $X$ and $Y$, respectively:
\begin{equation}
\begin{aligned}
& \frac{\partial^2 c_j}{\partial \boldsymbol{X}^2}=\xi \mu k_y\left(\ddot{y}\left(\tau_j\right) \frac{\partial \tau_j}{\partial \boldsymbol{X}} \frac{\partial \tau_j^T}{\partial \boldsymbol{X}}+\dot{y}\left(\tau_j\right) \frac{\partial^2 \tau_j}{\partial \boldsymbol{X}^2}\right) \\
& \quad-k_x\left(\dot{\boldsymbol{H}}_{-}\left(\tau_j\right) \frac{\partial \tau_j^T}{\partial \boldsymbol{X}}+\frac{\partial \tau_j}{\partial \boldsymbol{X}} \dot{\boldsymbol{H}}_{-}\left(\tau_j\right)+\ddot{x}\left(\tau_j\right) \frac{\partial \tau_j}{\partial \boldsymbol{X}} \frac{\partial \tau_j^T}{\partial \boldsymbol{X}}+\dot{x}\left(\tau_j\right) \frac{\partial^2 \tau_j}{\partial \boldsymbol{X}^2}\right)
\end{aligned}
\end{equation}
\begin{equation}
\begin{aligned}
& \frac{\partial^2 c_j}{\partial \boldsymbol{X} \partial \boldsymbol{Y}}=\xi \mu k_y\left(\ddot{y}\left(\tau_j\right) \frac{\partial \tau_j}{\partial \boldsymbol{X}} \frac{\partial \tau_j{ }^T}{\partial \boldsymbol{Y}}+\frac{\partial \tau_j}{\partial \boldsymbol{X}} \dot{\boldsymbol{H}}_{-}\left(\tau_j\right)^T+\dot{y}\left(\tau_j\right) \frac{\partial^2 \tau_j}{\partial \boldsymbol{X} \partial \boldsymbol{Y}}\right) \\
&\quad -k_x\left(\dot{\boldsymbol{H}}_{-}\left(\tau_j\right) \frac{\partial \tau_j{ }^T}{\partial \boldsymbol{Y}}+\ddot{x}\left(\tau_j\right) \frac{\partial \tau_j}{\partial \boldsymbol{X}} \frac{\partial \tau_j{ }^T}{\partial \boldsymbol{Y}}+\dot{x}\left(\tau_j\right) \frac{\partial^2 \tau_j}{\partial \boldsymbol{X} \partial \boldsymbol{Y}}\right)
\end{aligned}
\end{equation}
\begin{equation}
\begin{aligned}
& \frac{\partial^2 c_j}{\partial \boldsymbol{Y} \partial \boldsymbol{X}}=\xi \mu k_y\left(\dot{\boldsymbol{H}}_{-}\left(\tau_j\right) \frac{\partial \tau_j^T}{\partial \boldsymbol{X}}+\ddot{y}\left(\tau_j\right) \frac{\partial \tau_j}{\partial \boldsymbol{X}} \frac{\partial \tau_j^T}{\partial \boldsymbol{X}}+\dot{y}\left(\tau_j\right) \frac{\partial^2 \tau_j}{\partial \boldsymbol{Y} \partial \boldsymbol{X}}\right) \\ 
&\quad -k_x\left(\ddot{x}\left(\tau_j\right) \frac{\partial \tau_j}{\partial \boldsymbol{Y}} \frac{\partial \tau_j^T}{\partial \boldsymbol{X}}+\frac{\partial \tau_j}{\partial \boldsymbol{Y}} \dot{\boldsymbol{H}}_{-}\left(\tau_j\right)^T\right. 
\left.+\dot{x}\left(\tau_j\right) \frac{\partial^2 \tau_j}{\partial \boldsymbol{Y} \partial \boldsymbol{X}}\right)
\end{aligned}
\end{equation}
\begin{equation}
\begin{aligned}
& \frac{\partial^2 c_j}{\partial \boldsymbol{Y}^2}=\xi \mu k_y\left(\dot{\boldsymbol{H}}_{-}\left(\tau_j\right) \frac{\partial \tau_j^T}{\partial \boldsymbol{Y}}+\ddot{y}\left(\tau_j\right) \frac{\partial \tau_j}{\partial \boldsymbol{X}} \frac{\partial \tau_j^T}{\partial \boldsymbol{Y}}+\frac{\partial \tau_j}{\partial \boldsymbol{Y}} \dot{\boldsymbol{H}}_{-}\left(\tau_j\right)^T+\dot{y}\left(\tau_j\right) \frac{\partial^2 \tau_j}{\partial \boldsymbol{Y}^2}\right) \\
& -k_x\left(\ddot{x}\left(\tau_j\right) \frac{\partial \tau_j}{\partial \boldsymbol{Y}} \frac{\partial \tau_j^T}{\partial \boldsymbol{Y}}\right. 
\left.+\dot{x}\left(\tau_j\right) \frac{\partial^2 \tau_j}{\partial \boldsymbol{Y}^2}\right),
\end{aligned}
\end{equation}
where $\frac{\partial^2 \tau_j}{\partial \boldsymbol{X}^2}$, $\frac{\partial^2 \tau_j}{\partial \boldsymbol{X} \partial \boldsymbol{Y}}$, $\frac{\partial^2 \tau_j}{\partial \boldsymbol{Y} \partial \boldsymbol{X}}$, and $\frac{\partial^2 \tau_j}{\partial \boldsymbol{Y}^2}$ are obtained by taking the partial derivatives of the transition time, which is from either the slip or separation state to the stick state, with respect to $X$ and $Y$, respectively. Similarly, their values are determined by considering two separate cases.
When $\tau_j$ is the transition time from the slip state to the stick state,
\begin{equation}
\begin{aligned}
&\frac{\partial^2 \tau_j}{\partial \boldsymbol{X}^2}=\frac{-\xi k_x}{\xi k_x \ddot{x}\left(\tau_{\text {stick }}\right)-\mu k_y \ddot{y}\left(\tau_{\text {stick }}\right)} \ddot{\boldsymbol{H}}_{-}\left(\tau_{\text {stick }}\right) \frac{\partial \tau_j^T}{\partial \boldsymbol{X}}\\
&+\frac{\left(\xi k_x\right)^2 }{\left(\xi k_x \ddot{x}\left(\tau_{\text {stick }}\right)-\mu k_y \ddot{y}\left(\tau_{\text {stick }}\right)\right)^2}\dot{\boldsymbol{H}}_{-}\left(\tau_{\text {stick }}\right) \ddot{\boldsymbol{H}}_{-}\left(\tau_{\text {stick }}\right)^T \\
&+\frac{\left(\xi k_x\right)^2 \dddot{x}\left(\tau_{\text {stick }}\right) }{\left(\xi k_x \ddot{x}\left(\tau_{\text {stick }}\right)-\mu k_y \ddot{y}\left(\tau_{\text {stick }}\right)\right)^2} \dot{\boldsymbol{H}}_{-}\left(\tau_{\text {stick }}\right) \frac{\partial \tau_j^T}{\partial \boldsymbol{X}}\\
&-\frac{\left(\xi k_x \mu k_y\right) \dddot{y}\left(\tau_{\text {stick }}\right) }{\left(\xi k_x \ddot{x}\left(\tau_{\text {stick }}\right)-\mu k_y \ddot{y}\left(\tau_{\text {stick }}\right)\right)^2}\dot{\boldsymbol{H}}_{-}\left(\tau_{\text {stick }}\right) \frac{\partial \tau_j^T}{\partial \boldsymbol{X}}
\end{aligned}
\end{equation}
\begin{equation}
\begin{aligned}
&\frac{\partial^2 \tau_j}{\partial \boldsymbol{X} \partial \boldsymbol{Y}}=\frac{-\xi k_x}{\xi k_x \ddot{x}\left(\tau_{\text {stick }}\right)-\mu k_y \ddot{y}\left(\tau_{\text {stick }}\right)} \ddot{\boldsymbol{H}}_{-}\left(\tau_{\text {stick }}\right) \frac{\partial \tau_j^T}{\partial \boldsymbol{Y}}\\
&-\frac{\left(\xi k_x \mu k_y\right) }{\left(\xi k_x \ddot{x}\left(\tau_{\text {stick }}\right)-\mu k_y \ddot{y}\left(\tau_{\text {stick }}\right)\right)^2} \dot{\boldsymbol{H}}_{-}\left(\tau_{\text {stick }}\right) \ddot{\boldsymbol{H}}_{-}\left(\tau_{\text {stick }}\right)^T\\
&-\frac{\left(\xi k_x \mu k_y\right) \dddot{y}\left(\tau_{\text {stick }}\right) }{\left(\xi k_x \ddot{x}\left(\tau_{\text {stick }}\right)-\mu k_y \ddot{y}\left(\tau_{\text {stick }}\right)\right)^2}\dot{\boldsymbol{H}}_{-}\left(\tau_{\text {stick }}\right) \frac{\partial \tau_j^T}{\partial \boldsymbol{Y}}\\
&+\frac{\left(\xi k_x\right)^2 \dddot{x}\left(\tau_{\text {stick }}\right) }{\left(\xi k_x \ddot{x}\left(\tau_{\text {stick }}\right)-\mu k_y \ddot{y}\left(\tau_{\text {stick }}\right)\right)^2}\dot{\boldsymbol{H}}_{-}\left(\tau_{\text {stick }}\right) \frac{\partial \tau_j^T}{\partial \boldsymbol{Y}}
\end{aligned}
\end{equation}
\begin{equation}
\begin{aligned}
&\frac{\partial^2 \tau_j}{\partial \boldsymbol{Y} \partial \boldsymbol{X}}=  \frac{\mu k_y}{\xi k_x \ddot{x}\left(\tau_{\text {stick }}\right)-\mu k_y \ddot{y}\left(\tau_{\text {stick }}\right)} \ddot{\boldsymbol{H}}_{-}\left(\tau_{\text {stick }}\right) \frac{\partial \tau_j^T}{\partial \boldsymbol{X}} \\
& +\frac{\left(\mu k_y\right)^2 \dddot{y}\left(\tau_{\text {stick }}\right)}{\left(\xi k_x \ddot{x}\left(\tau_{\text {stick }}\right)-\mu k_y \ddot{y}\left(\tau_{\text {stick }}\right)\right)^2} \dot{\boldsymbol{H}}_{-}\left(\tau_{\text {stick }}\right) \frac{\partial \tau_j^T}{\partial \boldsymbol{X}} \\
& -\frac{\left(\xi k_x \mu k_y\right) \dddot{x}\left(\tau_{\text {stick }}\right)}{\left(\xi k_x \ddot{x}\left(\tau_{\text {stick }}\right)-\mu k_y \ddot{y}\left(\tau_{\text {stick }}\right)\right)^2} \dot{\boldsymbol{H}}_{-}\left(\tau_{\text {stick }}\right) \frac{\partial \tau_j^T}{\partial \boldsymbol{X}} \\
& -\frac{\left(\xi k_x \mu k_y\right)}{\left(\xi k_x \ddot{x}\left(\tau_{\text {stick }}\right)-\mu k_y \ddot{y}\left(\tau_{\text {stick }}\right)\right)^2} \dot{\boldsymbol{H}}_{-}\left(\tau_{\text {stick }}\right) \ddot{\boldsymbol{H}}_{-}\left(\tau_{\text {stick }}\right)^T
\end{aligned}
\end{equation}
\begin{equation}
\begin{gathered}
\frac{\partial^2 \tau_j}{\partial \boldsymbol{Y}^2}=\frac{\mu k_y}{\xi k_x \ddot{x}\left(\tau_{\text {stick }}\right)-\mu k_y \ddot{y}\left(\tau_{\text {stick }}\right)} \ddot{\boldsymbol{H}}_{-}\left(\tau_{\text {stick }}\right) \frac{\partial \tau_j^T}{\partial \boldsymbol{Y}} \\
+\frac{\left(\mu k_y\right)^2 \dddot{y}\left(\tau_{\text {stick }}\right)}{\left(\xi k_x \ddot{x}\left(\tau_{\text {stick }}\right)-\mu k_y \ddot{y}\left(\tau_{\text {stick }}\right)\right)^2} \dot{\boldsymbol{H}}_{-}\left(\tau_{\text {stick }}\right) \frac{\partial \tau_j^T}{\partial \boldsymbol{Y}} \\
-\frac{\left(\xi k_x \mu k_y\right) \dddot{x}\left(\tau_{\text {stick }}\right)}{\left(\xi k_x \ddot{x}\left(\tau_{\text {stick }}\right)-\mu k_y \ddot{y}\left(\tau_{\text {stick }}\right)\right)^2} \dot{\boldsymbol{H}}_{-}\left(\tau_{\text {stick }}\right) \frac{\partial \tau_j^T}{\partial \boldsymbol{Y}} \\
+\frac{\left(\mu k_y\right)^2}{\left(\xi k_x \ddot{x}\left(\tau_{\text {stick }}\right)-\mu k_y \ddot{y}\left(\tau_{\text {stick }}\right)\right)^2} \dot{\boldsymbol{H}}_{-}\left(\tau_{\text {stick }}\right) \ddot{\boldsymbol{H}}_{-}\left(\tau_{\text {stick }}\right)^T.
\end{gathered}
\end{equation}
When $\tau_j$ is the transition time from the separation state to the stick state,
\begin{equation}
\frac{\partial^2 \tau_j}{\partial \boldsymbol{X}^2}= \boldsymbol{0}
\end{equation}
\begin{equation}
\frac{\partial^2 \tau_j}{\partial \boldsymbol{X} \boldsymbol{Y}}= \boldsymbol{0}
\end{equation}
\begin{equation}
\frac{\partial^2 \tau_j}{\partial \boldsymbol{Y} \partial \boldsymbol{X}}=\frac{1}{-\dot{y}\left(\tau_{\text {stick }}\right)} \dot{\boldsymbol{H}}_{-}\left(\tau_{\text {stick }}\right) \frac{\partial \tau_j^T}{\partial \boldsymbol{X}}+\frac{\ddot{y}\left(\tau_{\text {stick }}\right)}{\left(\dot{y}\left(\tau_{\text {stick }}\right)\right)^2} \boldsymbol{H}_{-}\left(\tau_{\text {stick }}\right) \frac{\partial \tau_j^T}{\partial \boldsymbol{X}}
\end{equation}
\begin{equation}
\begin{gathered}
\frac{\partial^2 \tau_j}{\partial \boldsymbol{Y}^2}=\frac{1}{-\dot{y}\left(\tau_{\text {stick }}\right)} \dot{\boldsymbol{H}}_{-}\left(\tau_{\text {stick }}\right) \frac{\partial \tau_j^T}{\partial \boldsymbol{Y}}+\frac{\ddot{y}\left(\tau_{\text {stick }}\right)}{\left(\dot{y}\left(\tau_{\text {stick }}\right)\right)^2} \boldsymbol{H}_{-}\left(\tau_{\text {stick }}\right) \frac{\partial \tau_j^T}{\partial \boldsymbol{Y}} \\
+\frac{1}{\left(\dot{y}\left(\tau_{\text {stick }}\right)\right)^2} \boldsymbol{H}_{-}\left(\tau_{\text {stick }}\right) \dot{\boldsymbol{H}}_{-}\left(\tau_{\text {stick }}\right)^T.
\end{gathered}
\end{equation}

Following the same logic, the $2N_h+i$-th Hessian matrix, obtained by differentiating the i-th row $\boldsymbol{K}_f(2N_h+i,:)$ of the Jacobian matrix $\boldsymbol{K}_f$, is expressed as
\begin{equation}
\operatorname{Hes}_{2 N h+i}=\left[\frac{\partial \boldsymbol{K}_f(2 N h+i,:)^T}{\partial \boldsymbol{X}}, \frac{\partial \boldsymbol{K}_f(2 N h+i,:)^T}{\partial \boldsymbol{Y}}\right]=\sum_{j=1}^{n_\tau}\left[\begin{array}{cc}
\mathbf{0} & \mathbf{0} \\
\mathbf{0} & \frac{\partial \frac{\partial \boldsymbol{J}_y^{(j)}}{\partial \boldsymbol{Y}}(i,:)^T}{\partial \boldsymbol{Y}}
\end{array}\right],
\end{equation}

where
\begin{equation}
\frac{\partial \frac{\partial \boldsymbol{J}_y^{(j)}}{\partial \boldsymbol{Y}}(i,:)^T}{\partial \boldsymbol{Y}}= \begin{cases} k_y\left(\frac{\partial \boldsymbol{W}_j(i,:)^T}{\partial \tau_{j+1}} \frac{\partial \tau_{j+1}}{\partial \boldsymbol{Y}}-\frac{\partial \boldsymbol{W}_j(i,:)^T}{\partial \tau_j} \frac{\partial \tau_j}{\partial \boldsymbol{Y}}\right) & \text { contact } \\ \mathbf{0} & \text { separation }\end{cases}.
\end{equation}
After obtaining the Hessian Tensor of the contact element, the second derivatives of Equation (21) with respect to the variables $\alpha$ and $\omega$, as well as the second derivative of $A_m$ with respect to $Q$, must be computed, which are given as

\begin{equation}
\left\{\begin{array}{c}
\frac{\partial^2 A m}{\partial Q_{k,i}{ }^2}=\frac{\|\boldsymbol{Q_k}\|^2-Q_{k,i}{ }^2}{\|\boldsymbol{Q_k}\|^3} \\
\frac{\partial^2 A m}{\partial Q_{k,i} \partial Q_{k,j}}=-\frac{Q_{k,i} Q_{k,j}}{\|\boldsymbol{Q_k}\|^3} \quad i \neq j
\end{array}\right.
\end{equation}

Here, $Q_{k,i}$ and $Q_{k,j}$ are the i-th and j-th components of vector $\boldsymbol{Q_k}$, respectively. Additionally, 
\begin{equation}
\begin{gathered}
\frac{\partial\left(\frac{\partial \boldsymbol{R}}{\partial \boldsymbol{Q}}\right)}{\partial \omega}=\frac{\partial D(\omega)}{\partial \omega},\quad \frac{\partial\left(\frac{\partial \boldsymbol{R}}{\partial \boldsymbol{Q}}\right)}{\partial \alpha}=\mathbf{0} \\
\left(\frac{\partial\left(\frac{\partial \boldsymbol{R}}{\partial \omega}\right)}{\partial \omega}\right)=\frac{\partial^2 D(\omega)}{\partial \omega^2},\quad\left(\frac{\partial\left(\frac{\partial \boldsymbol{R}}{\partial \omega}\right)}{\partial \alpha}\right)=\mathbf{0} \\
\left(\frac{\partial\left(\frac{\partial \boldsymbol{R}}{\partial \alpha}\right)}{\partial \omega}\right)=\mathbf{0}, \quad\left(\frac{\partial\left(\frac{\partial \boldsymbol{R}}{\partial \alpha}\right)}{\partial \alpha}\right)=\mathbf{0}.
\end{gathered}
\end{equation}

These partial derivatives with respect to the variables $(\boldsymbol{Q}, \omega, \alpha)$, together with the formulated Hessian Tensor of contact elements, constitute the Jacobian matrix of Equation (22). All the aforementioned equations are analytical, and the Newton--Raphson iterative method can be used to rapidly solve the equations for the resonance backbone curve of a system with friction.
\section{Numerical validation} \label{sec:Numerical}
\subsection{Cantilever beam with contact element} \label{ssec:instance}
We considered a slender cantilever beam in contact with a fixed pad under a constant normal load. The beam was subjected to sinusoidal excitation, and the locations of the contact and applied forces are illustrated in Figure 4.
\begin{figure}
  \centering
  \includegraphics[width=0.8\textwidth]{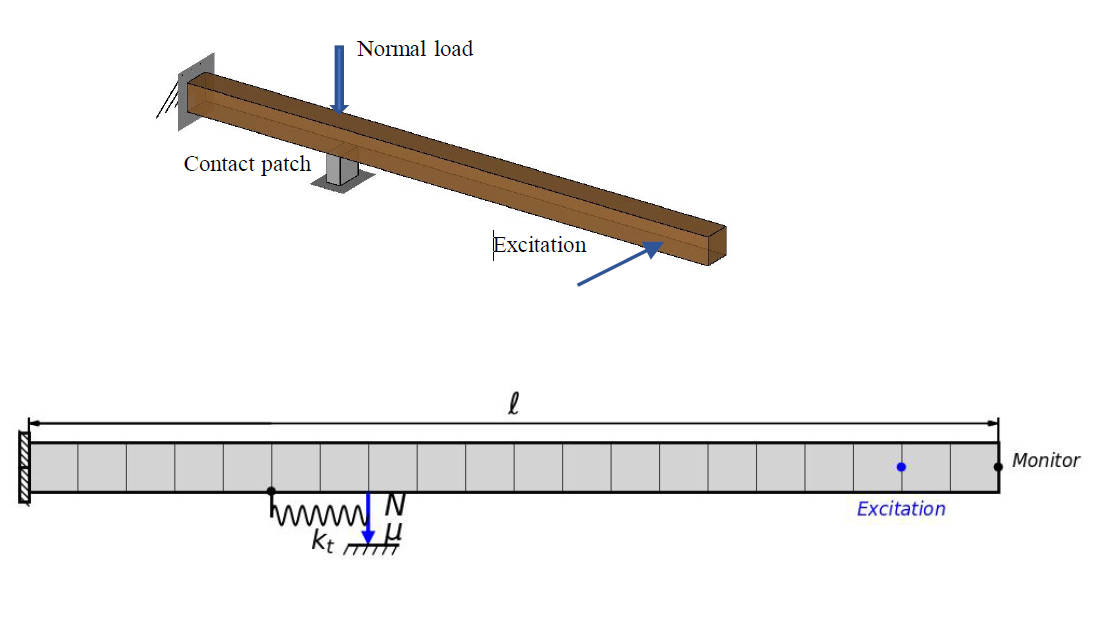}
  \caption{Schematic of the cantilever beam in contact with a fixed pad}
\end{figure}

The material properties of the beam model and parameters of the contact elements are listed in Table 1.
\begin{table}[t]
\caption[Table]{Material properties of the beam model and the parameters of the contact element}\label{tab:1}
\centering{%
\begin{tabular}{llr}
\toprule
Parameter & Value \\
\midrule
Young' modulus &210 [Gpa] \\
Density & 7830 [kg/$m^3$] \\
Beam length $\times$ width $\times$ height & 200 $\times$ 10 $\times$ 10 [mm] \\
Number of elements & 20 \\
Contact stiffness($k_t$ and $k_n$) & $2*10^8$ [N/m] \\
Friction coefficient($\mu$) & 0.3 \\
Initial normal load ($N_0$) & 40 [N] \\
\bottomrule
\end{tabular}
}
\end{table}

The beam was modelled using classical Euler--Bernoulli elements. The model consisted of 20 elements and 21 nodes (excluding the fixed node at the clamp). Each node comprised two DOFs: the transverse displacement and rotation angle. The locations of the contact, excitation, and monitoring points are shown in Figure 4.

The forced vibration response equation for the model can be expressed in the form of Equation (22). First, by retaining only the fundamental Fourier harmonic, the parameter continuation method developed in this study was employed to solve for the resonance backbone curve. The results are shown in Figure 5(a), where the black line represents the initial FRC at the starting excitation level, corresponding to the first stage of the continuation process. The red dots represent the second stage, where the design variables are constant while continuation is performed on the Lagrange multipliers. Finally, the magenta curve denotes the resulting resonance backbone curve obtained from the third stage of the continuation.

Figure 5 presents the excitation load and response amplitude axes on a logarithmic scale. The projected curves in the planes (amplitude and excitation) enable the observation of the energy-dependent nature of the resonance response and frequency. The system behaved linearly at low load levels, as the contact interface was fully stuck. Conversely, the system exhibited linear characteristics at high excitation levels as the contact interface approached a fully free state. In both the limiting cases, the system displayed the features of a linear system.
\begin{figure}[htbp] 
    \centering
    \begin{subfigure}{0.5\textwidth} 
        \centering 
        \includegraphics[width=\linewidth]{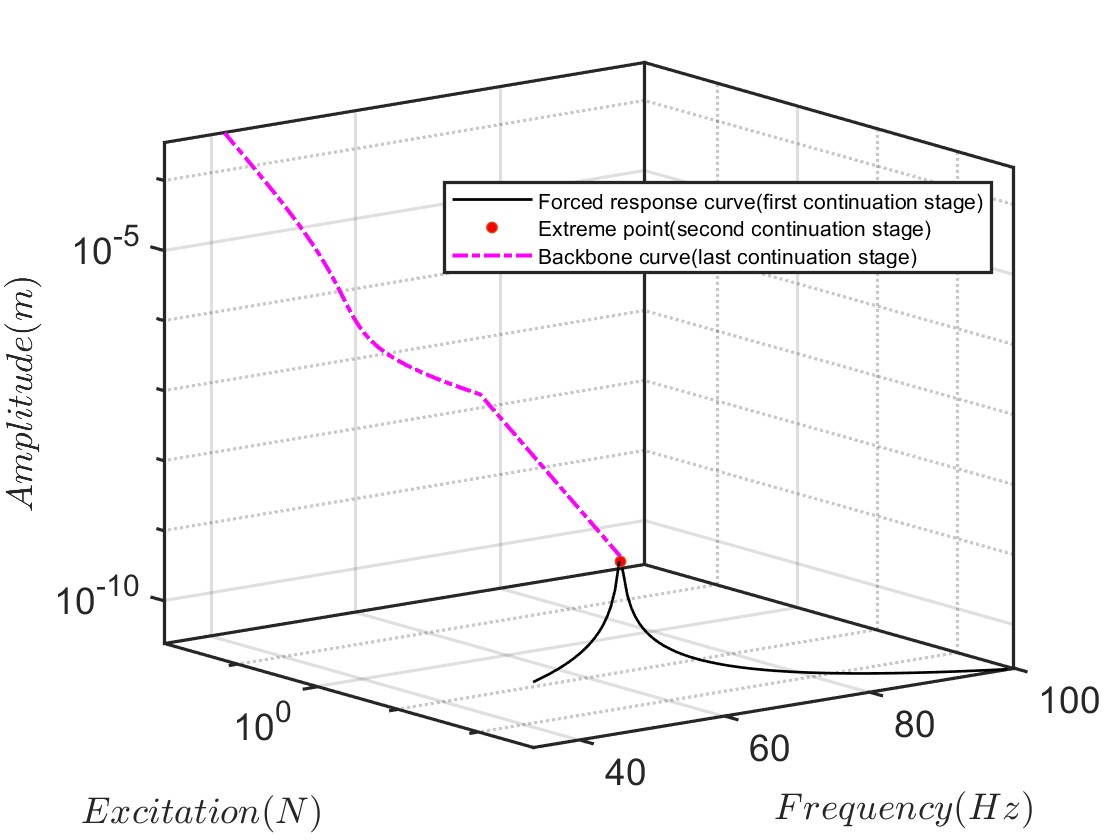}
        \caption{Three stages of the continuation process and their respective curves}
        \label{fig:sub1}
    \end{subfigure}
    \\ 
    \begin{subfigure}{0.48\textwidth}
        \centering
        \includegraphics[width=\linewidth]{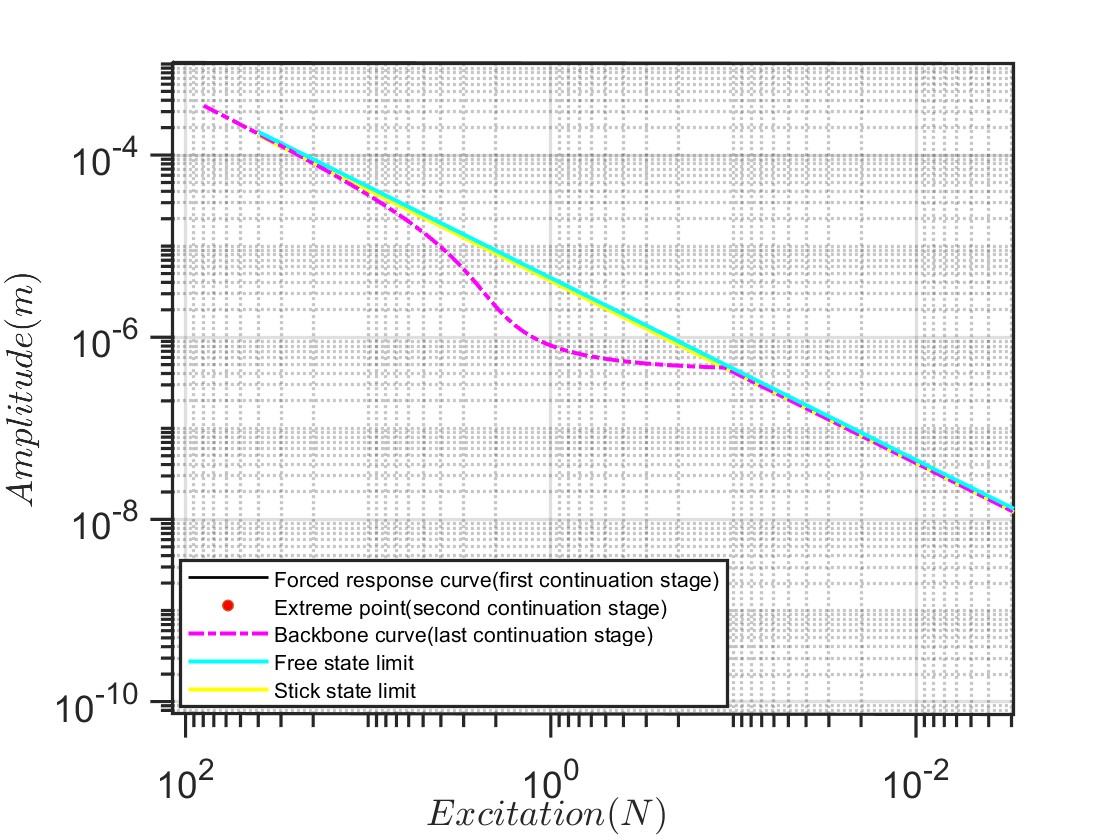}
        \caption{Excitation--amplitude plane}
        \label{fig:sub2}
    \end{subfigure}
    \hfill 
    \begin{subfigure}{0.48\textwidth}
        \centering
        \includegraphics[width=\linewidth]{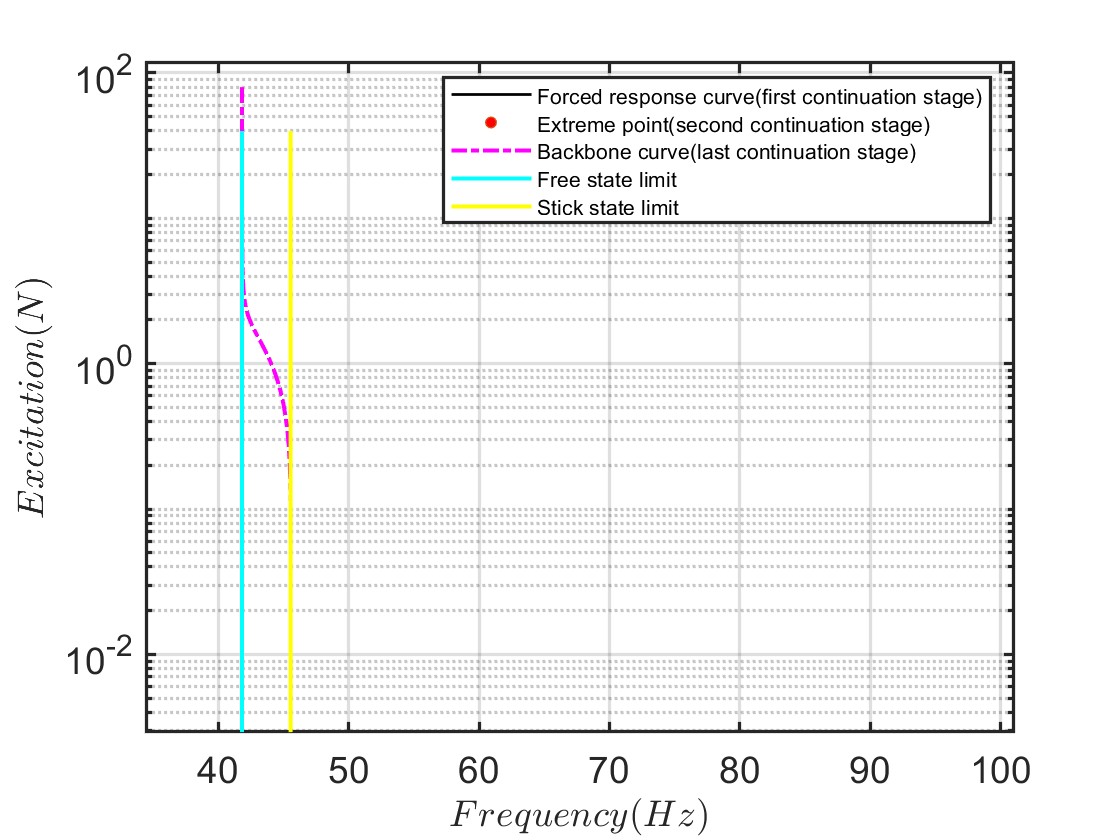}
        \caption{Frequency--excitation plane}
        \label{fig:sub3}
    \end{subfigure}
    \caption{Backbone curve obtained via continuation process}
    \label{fig:overall}
\end{figure}

The backbone curve accurately captures the resonance peaks and their corresponding frequencies at different excitation levels. Figure 6 shows these results agree well with the resonance points obtained from direct forced response computations. For comparison, the resonance backbone curve was also computed using the dNNMs and the phase resonance methods, as shown in Figure 6. 

The results indicated that all three methods yielded consistent outcomes for most considered ranges. However, the magnified view in Figure 6 reveals a key advantage: only the proposed continuation method can effectively capture the sharp local changes that occur as the contact interface transitions from a fully stuck state to a partial-slip state.

\begin{figure}
  \centering
  \includegraphics[width=0.9\textwidth]{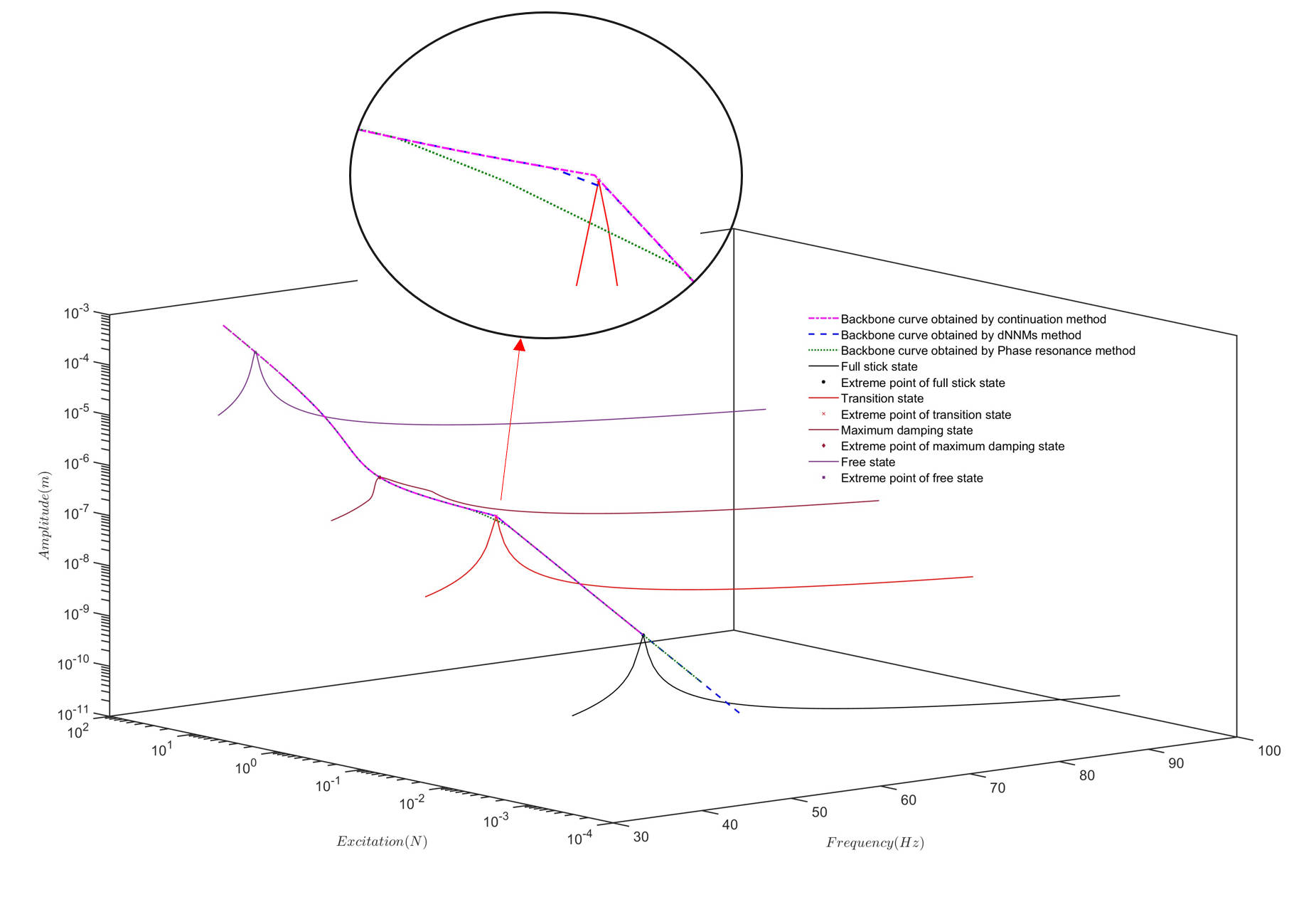}
  \caption{Resonance backbone curve and the FRCs under different contact states}
\end{figure}

Correspondingly, the resonant mode shapes for the four resonance points (representing four different contact states) on the backbone curve in Figure 6 can be plotted, as shown in Figure 7.
\begin{figure}
  \centering
  \begin{subfigure}[h]{0.45\textwidth}
    \includegraphics[width=\textwidth]{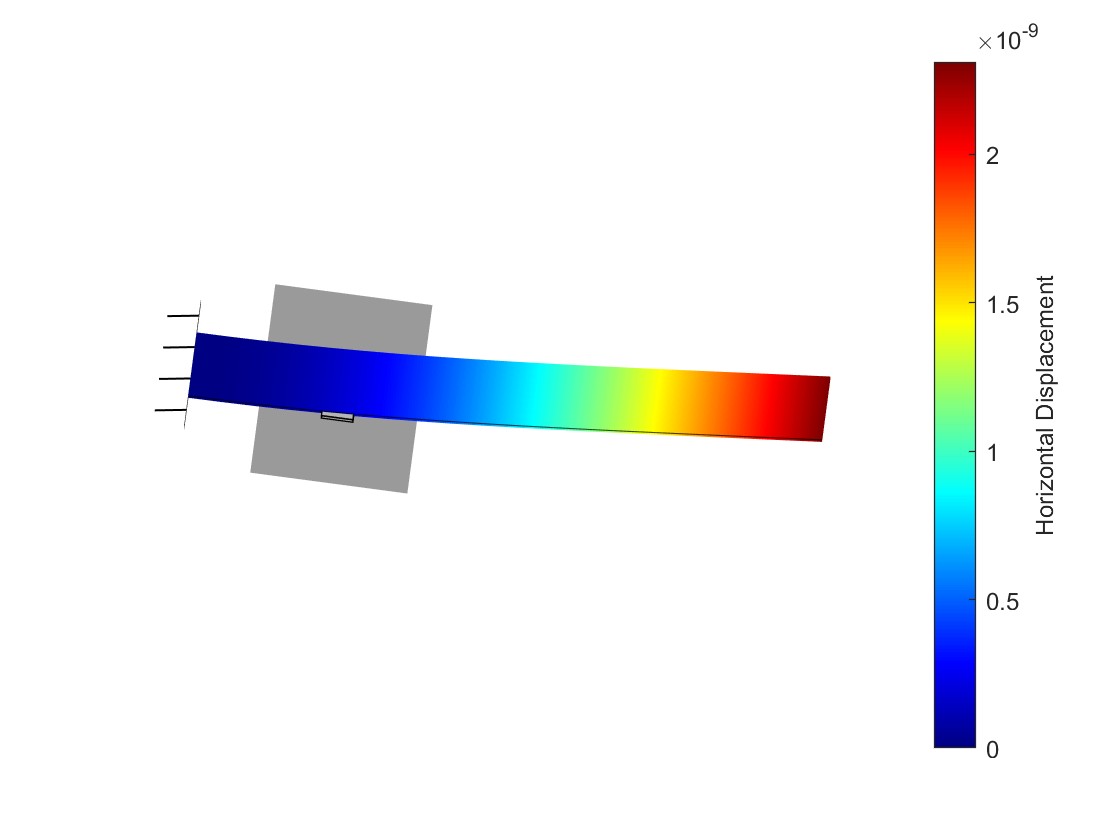}
    \caption{Fully stuck state}
  \end{subfigure}
  \hfill
  \begin{subfigure}[h]{0.45\textwidth}
    \includegraphics[width=\textwidth]{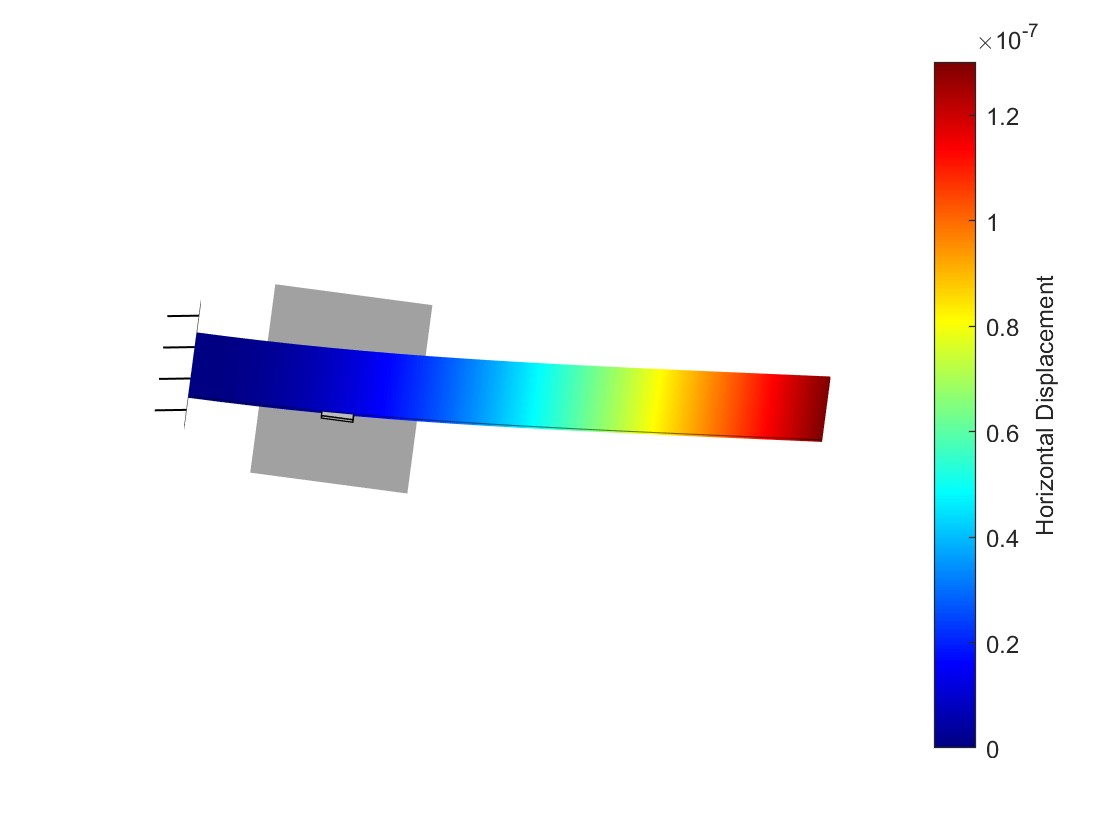}
  \caption{Transition state}  
  \end{subfigure}
  
  \begin{subfigure}[h]{0.45\textwidth}
    \includegraphics[width=\textwidth]{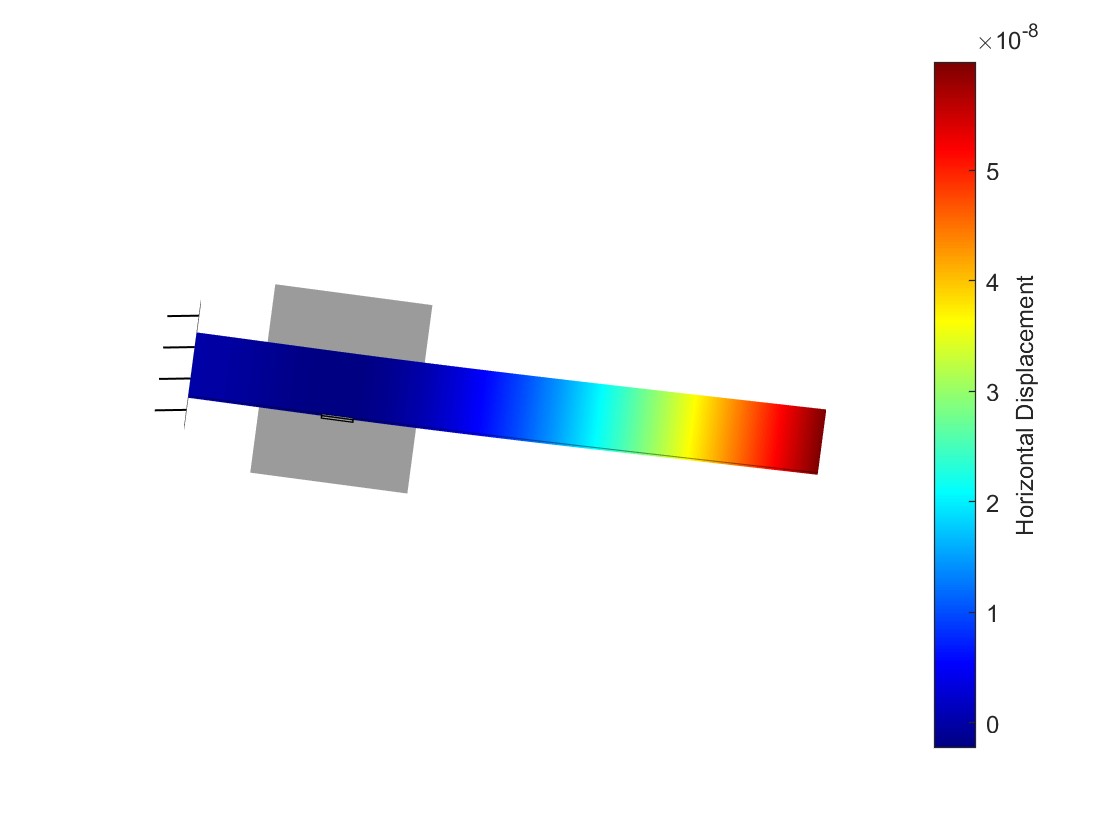}
  \caption{Maximum state}  
  \end{subfigure}
  \hfill
  \begin{subfigure}[h]{0.45\textwidth}
    \includegraphics[width=\textwidth]{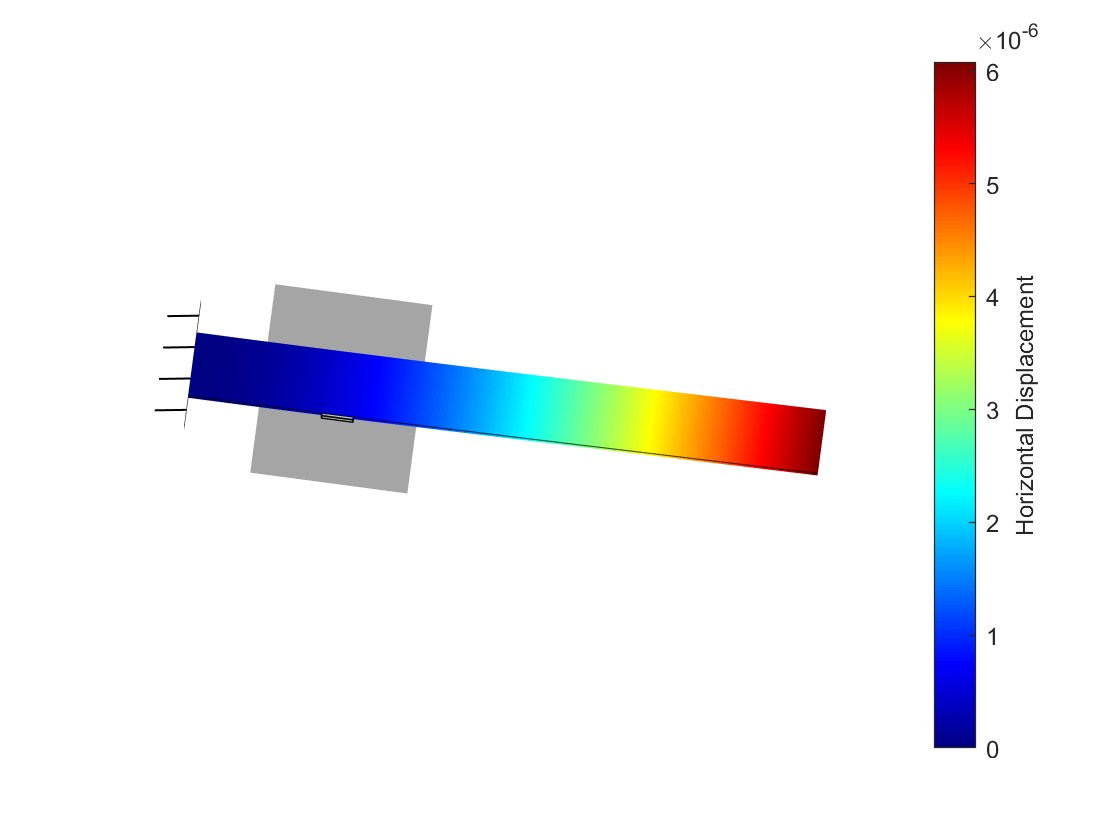}
  \caption{Free state}  
  \end{subfigure}
  \caption{Four initial normal pressure distribution scenarios on the contact patch}
\end{figure}

The colour map and deformation in the mode shape plots indicate that as the vibration energy increases, the relative displacement amplitude at the contact location first decreases and then increases. As the system approached a fully free state, the mode shape converged to the first linear mode of the cantilever beam. This behaviour is consistent with the trends commonly observed in the vibration analysis of turbine blades equipped with dry friction dampers.

A comparison between the results obtained using the first- and third-order Fourier harmonics is presented in Figure 7. The magnified view clearly shows that the proposed parameter continuation method identifies the resonance peak more accurately than the dNNMs and phase resonance methods. Notably, when a larger number of harmonics is retained, the dNNMs and phase resonance methods fail to capture the subtle changes caused by higher-order harmonics. In contrast, the continuation method can still distinguish between the solutions obtained with different harmonic orders. This distinction arises because the continuation method locates the resonance peak as a true extremum point on the response surface in a mathematical sense, whereas the other two methods provide only an approximate solution based on the phase-lag criterion. A further comparison, along with the limitations of the dNNMs and phase resonance methods, is presented in the next subsection, using the example of a bladed disk assembly with a dry friction damper.

\begin{figure}[htbp] 
    \centering
    \begin{subfigure}{0.77\textwidth} 
        \centering 
        \includegraphics[width=\linewidth]{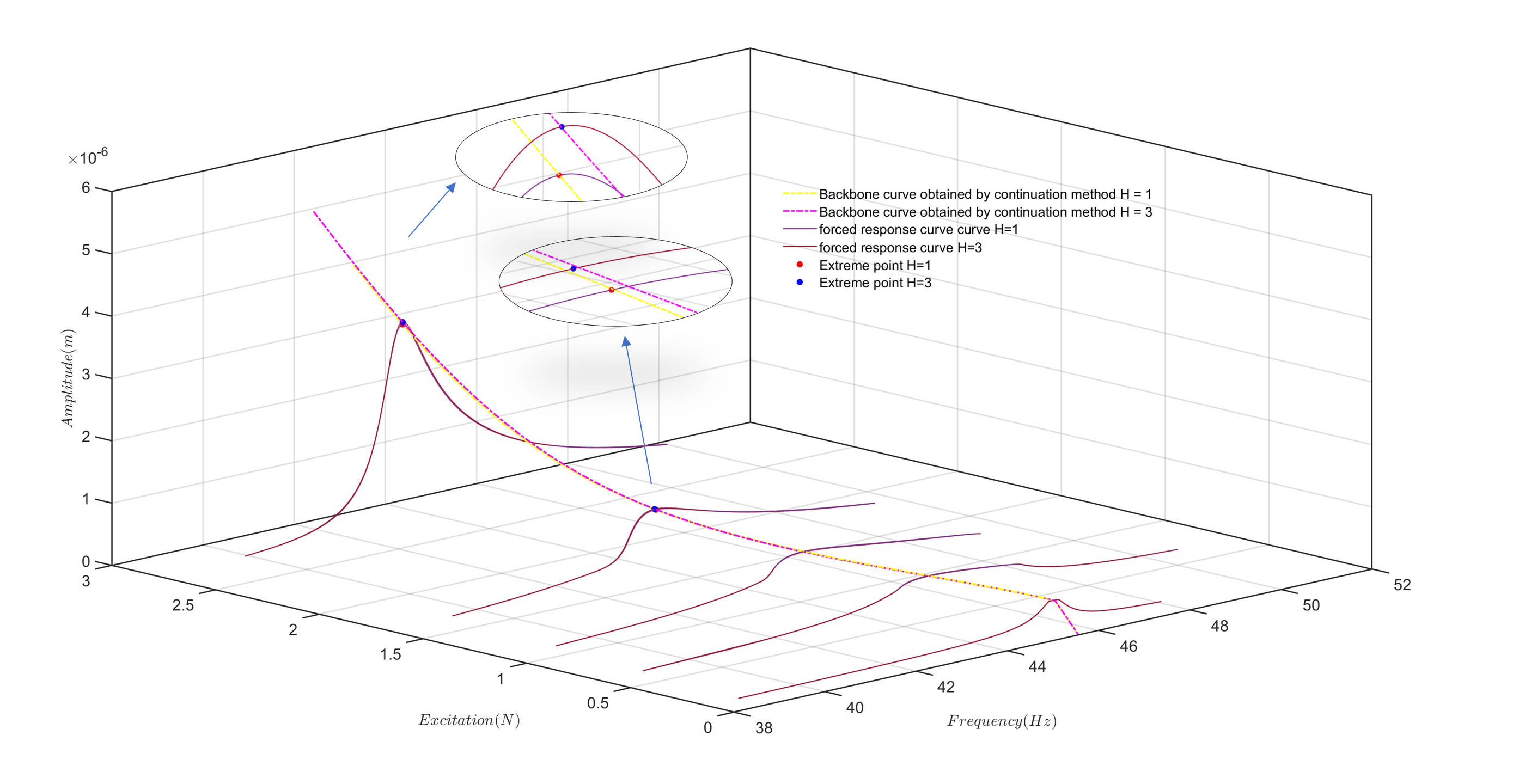}
        \caption{First- and third-order parameter continuation results with the forced response}
        \label{fig:sub1}
    \end{subfigure}
    \\ 
    \begin{subfigure}{0.77\textwidth}
        \centering
        \includegraphics[width=\linewidth]{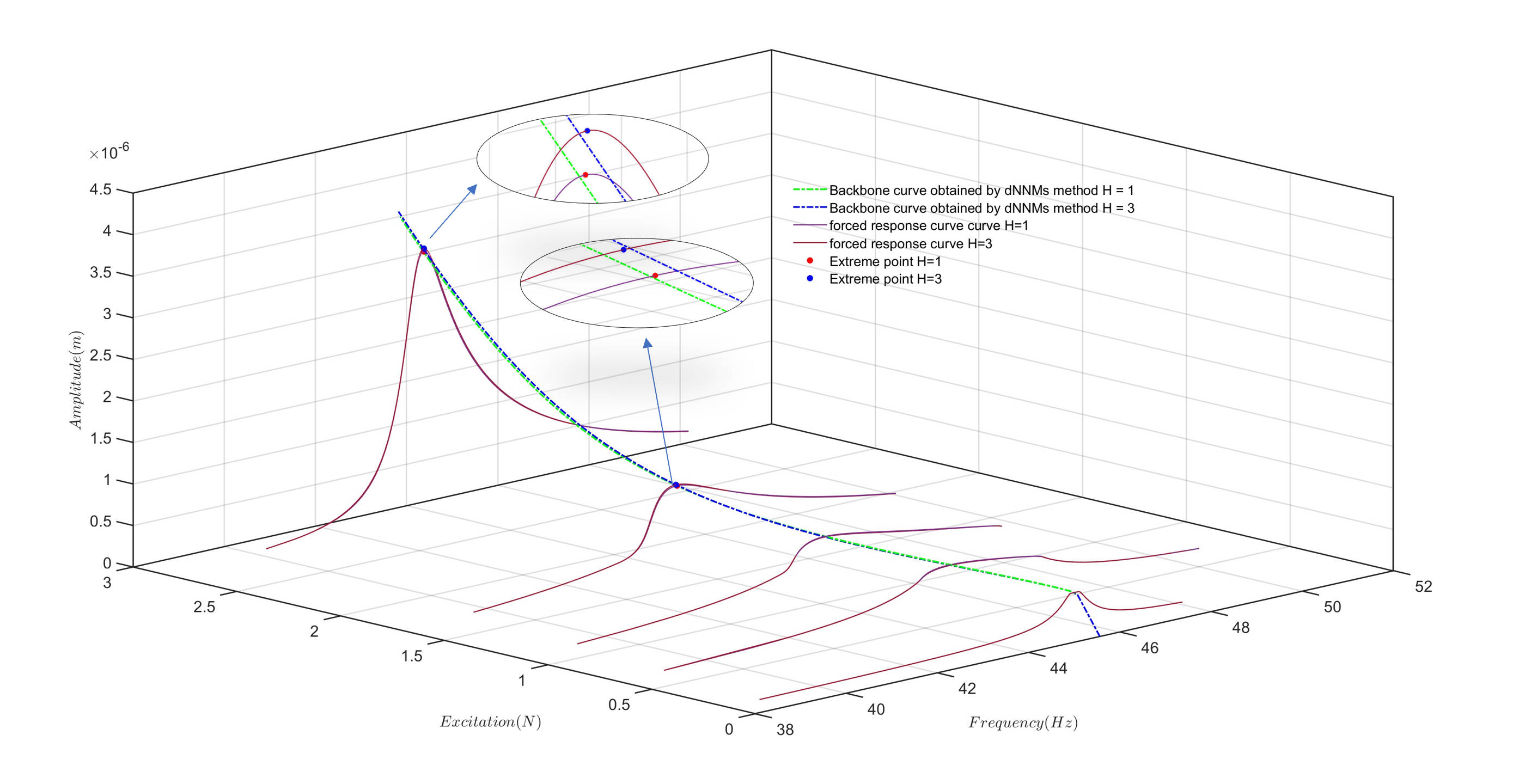}
        \caption{First- and third-order dNNMs results with the forced response}
        \label{fig:sub2}
    \end{subfigure}
    \hfill 
    \begin{subfigure}{0.77\textwidth}
        \centering
        \includegraphics[width=\linewidth]{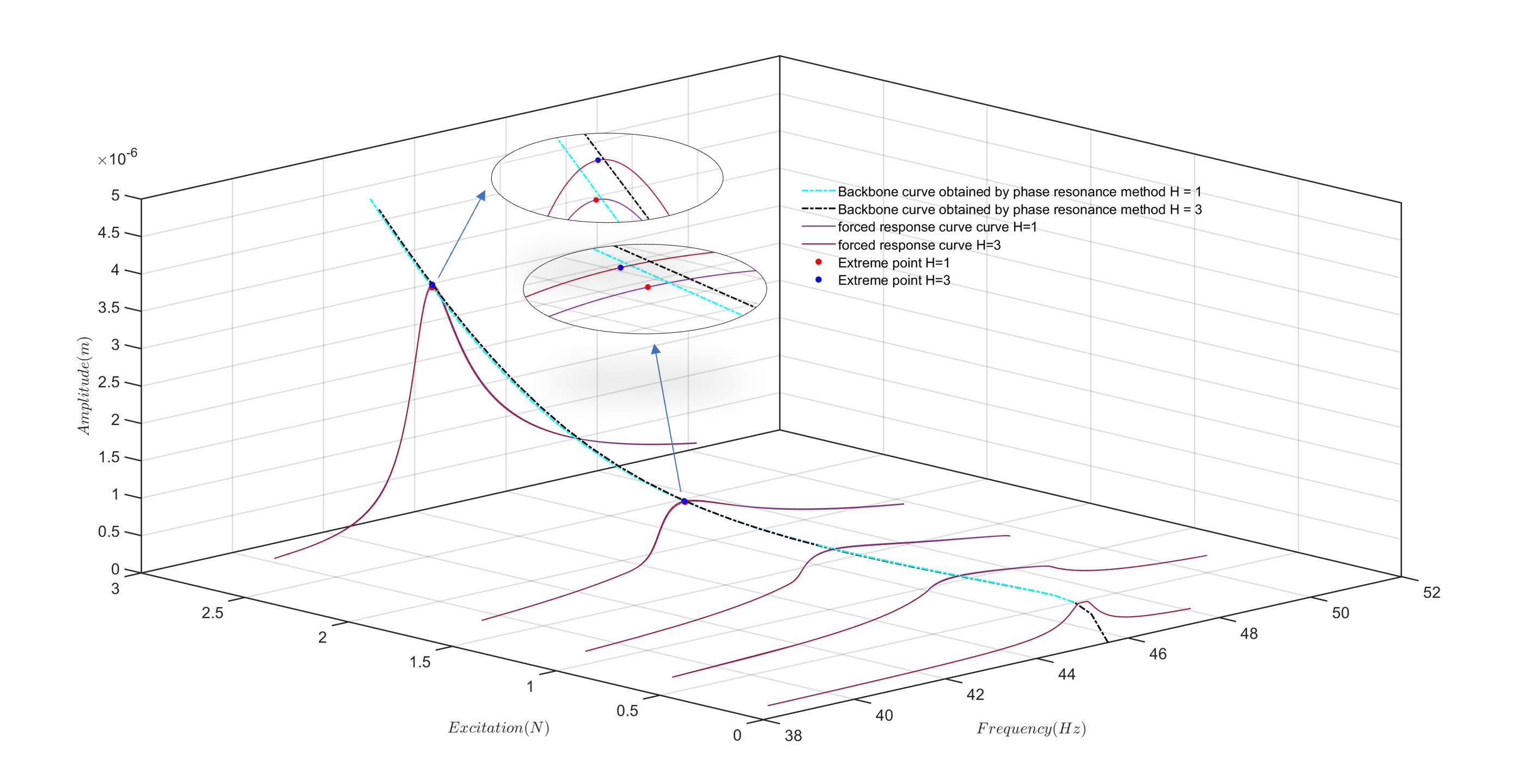}
        \caption{Frequency--excitation plane}
        \label{fig:sub3}
    \end{subfigure}
    \caption{First- and third-order phase resonance results with the forced response}
    \label{fig:overall}
\end{figure}

\subsection{Lumped-parameter model of blade--damper--blade} \label{ssec:instance}
A simplified lumped-parameter model for a blade--damper--blade system adapted from \cite{(26)Yang2019} is illustrated in Figure 9. The model parameters are listed in Table 1.
\begin{figure}
  \centering
  \includegraphics[width=0.8\textwidth]{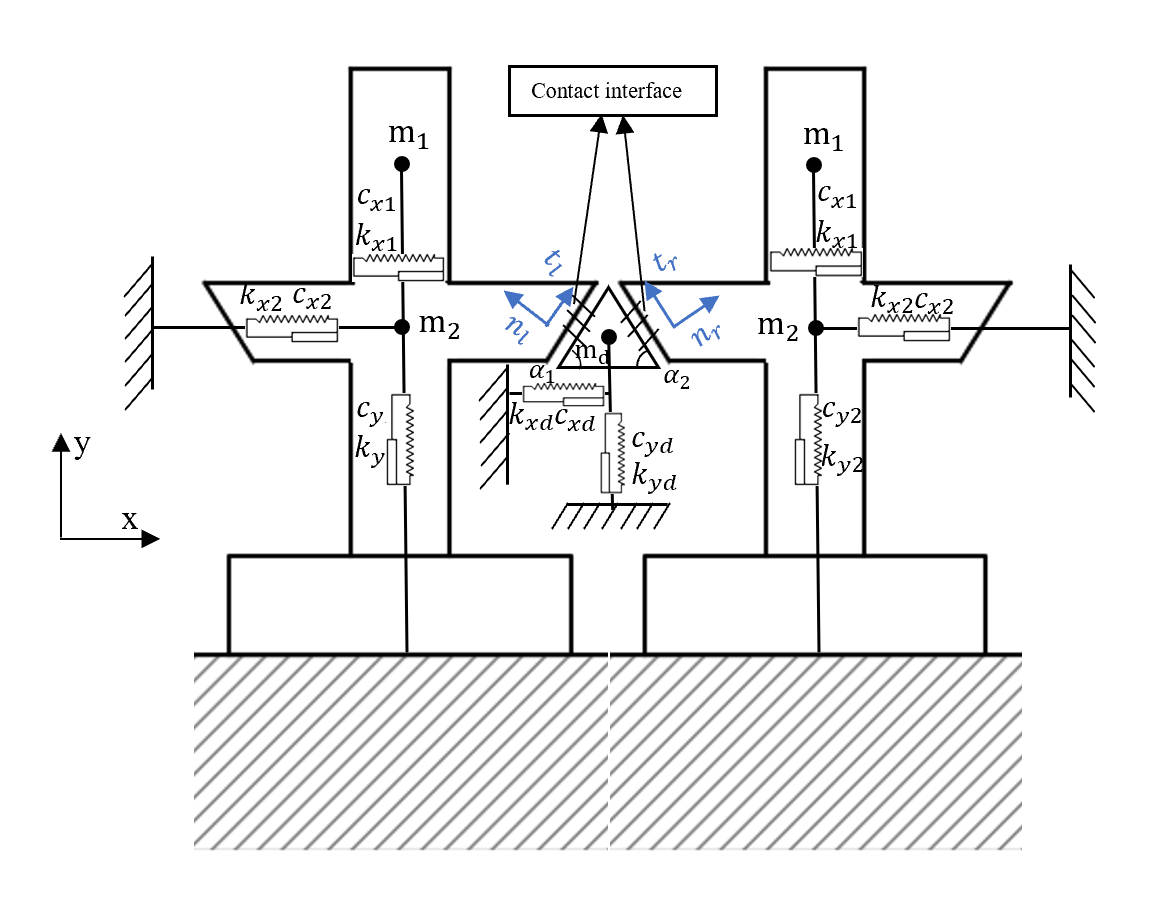}
  \caption{Lumped-parameter model of blades with an underplatform damper}
\end{figure}
\begin{table}[t]
\caption[Table]{Parameters of the lumped-parameter model}\label{tab:1}
\centering{%
\begin{tabular}{llr}
\toprule
Parameter & Value \\
\midrule
$k_x,k_y,k_{x1}$ & $3.95*10^5$ [N/m] \\
$k_{xd}$ & $2*10^5$ [N/m] \\
$k_{yd}$ & $1*10^5$ [N/m] \\
Contact stiffness($k_t$ and $k_n$) & $1*10^5$ [N/m] \\
$m_{1},m_{2}$ & $0.1$ [kg] \\
$m_{d}$ & $0.0044$ [kg] \\
$\alpha_{1},\alpha_{2}$ & $\frac{\pi}6 $ [rad] \\
$\beta$ damping factor & 0.000015 \\
Friction coefficient($\mu$) & 0.3 \\
Initial load on left and right interfaces ($N_0$) & 50 [N] \\
\bottomrule
\end{tabular}
}
\end{table}
As shown in Fig. 9, the blade and shroud are each represented by a point mass. The two masses are connected by a spring--damper element in the x-direction. The DOFs for the shroud and damper mass are connected to the ground via springs and dampers. The resulting equations of motion for the system can be expressed as follows:
\begin{equation}
\left\{\begin{array}{l}
m_1 \ddot{x}_1+c_{x 1}\left(\dot{x}_1-\dot{x}_2\right)+k_{x 1}\left(x_1-x_2\right)=0 \\
m_2 \ddot{x}_2+c_{x 1}\left(\dot{x}_2-\dot{x}_1\right)+c_x \dot{x}_2+k_{x 1}\left(x_2-x_1\right)+k_x x_2=f+F n l_{x l} \\
m_2 \ddot{y}_3+c_y \dot{y}_3+k_y y_3=F n l_{y 1} \\
m_{\mathrm{d}} \ddot{x}_4+c_{x \mathrm{~d}} \dot{x}_4+k_{x \mathrm{~d}} x_4=-F n l_{x l}-F n l_{x r} \\
m_{\mathrm{d}} \ddot{y}_5+c_{y \mathrm{~d}} \dot{y}_5+k_{y \mathrm{~d}} y_5=-F n l_{y l}-F n l_{y r} \\
m_2 \ddot{x}_6+c_{x 1}\left(\dot{x}_6-\dot{x}_8\right)+c_x \dot{x}_6+k_{x 1}\left(x_6-x_8\right)+k_x x_6=F n l_{x r} \\
m_2 \ddot{y}_7+c_y \dot{y}_7+k_y y_7=F n l_{y r} \\
m_1 \ddot{x}_8+c_{x 1}\left(\dot{x}_8-\dot{x}_6\right)+k_{x 1}\left(x_8-x_6\right)=0
\end{array}\right.
\end{equation}
where $f$ represents the excitation force applied to the blade, and $F n l_{x l}, F n l_{y l}, F n l_{x r}$ and $F n l_{x r}$ represent the contact forces in the x- and y-directions on the left and right contact interfaces of the shroud, respectively. These contact forces are obtained through coordinate transformation, as follows:
\begin{equation}
\begin{aligned}
& {\left[\begin{array}{l}
F n l_{x l} \\
F n l_{y l}
\end{array}\right]=\left[\begin{array}{cc}
\cos \alpha_1 & \sin \alpha_1 \\
\sin \alpha_1 & -\cos \alpha_1
\end{array}\right]\left[\begin{array}{l}
T_l \\
N_l
\end{array}\right]} \\
& {\left[\begin{array}{l}
F n l_{x r} \\
F n l_{x r}
\end{array}\right]=\left[\begin{array}{cc}
-\cos \alpha_2 & \sin \alpha_2 \\
-\sin \alpha_2 & -\cos \alpha_2
\end{array}\right]\left[\begin{array}{c}
T_r \\
N_r
\end{array}\right]}.
\end{aligned}
\end{equation}
In contrast to the beam model discussed in the previous subsection, the current lumped-parameter model enables consideration of the influence of the relative normal displacement on the normal contact force. The normal contact force N and tangential contact force T between the damper and shroud were calculated using the friction contact model described in Section 4.

When the initial normal load between the two contact surfaces was held constant at 50 N, single-point excitation was applied to the right-blade mass in the x-direction. The damping in the system was modelled as stiffness-proportional damping with a damping factor $\beta$ of 0.000015. The response was monitored at the same location (i.e., in the x-direction of the right-blade mass). Figure 10 presents the resonance backbone curves at different excitation levels, as computed via the parameter continuation, dNNMs, and phase resonance methods.
\begin{figure}
  \centering
  \includegraphics[width=0.98\textwidth]{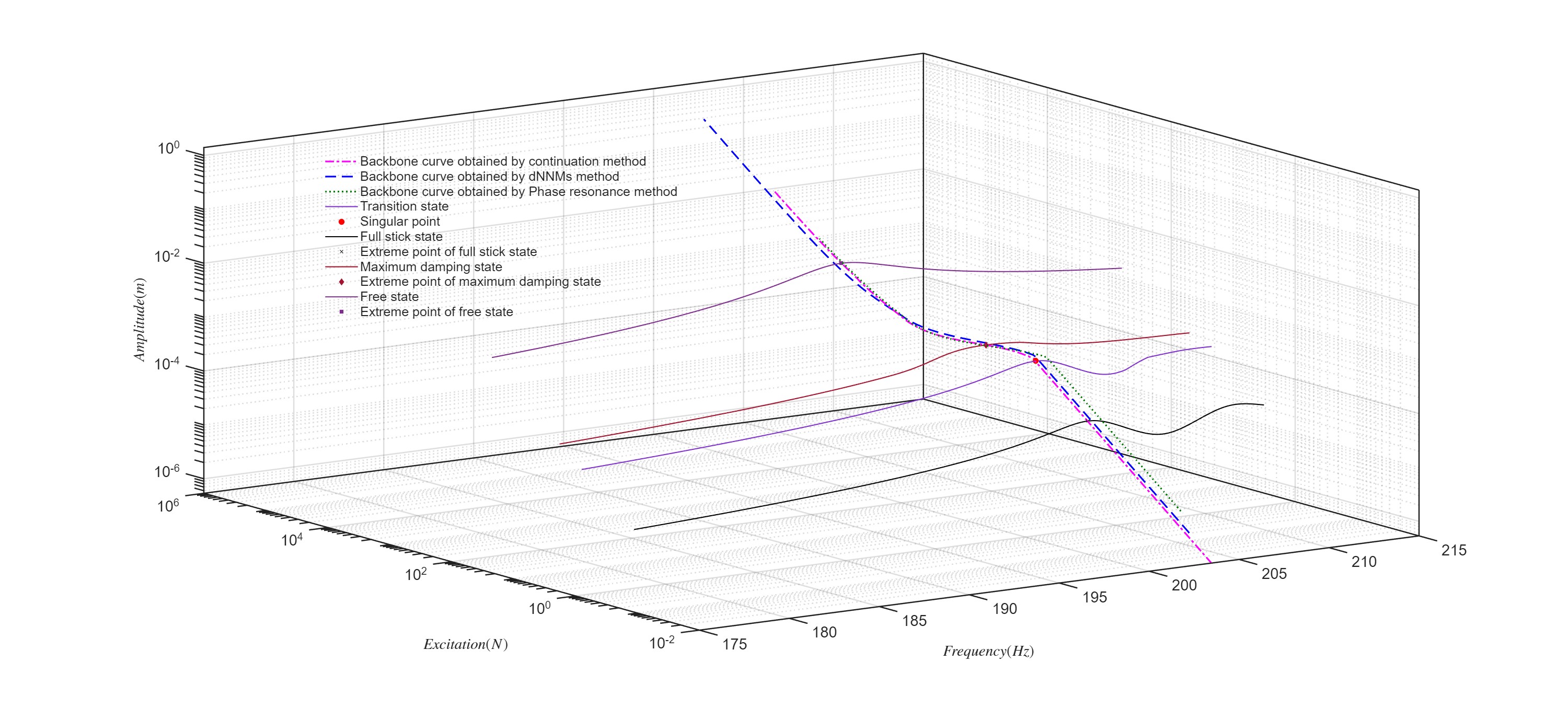}
  \caption{Resonance backbone curves of the lumped-parameter model obtained via different methods}
\end{figure}

As shown in the figure, the parameter continuation method accurately traces the resonance peaks of the forced response. Additionally, it is the only one capable of identifying the numerical singularity among the three methods. This singular point corresponds precisely to the contact transition between the damper and the right blade from a fully stuck state to the onset of slip. The continuation of the backbone curve past this singularity follows the procedure detailed in Section 3.2. In this case, the relative advantage of the parameter continuation method lies in the fact that the phase-lag criterion, which is used by the dNNMs and phase resonance methods to approximate resonance, is not applicable.

First, we consider the case where the contact interface is fully stuck, rendering the system linear. This linear system possesses two closely spaced bending modes: in-phase and out-of-phase, as shown in Figure 11. Owing to the modal coupling introduced by the inherent damping of the system, the resonance peaks for each blade are bifurcated, occurring at distinct frequencies. 

This phenomenon can be illustrated by plotting the FRCs of the blade-tip displacements in the x-direction. As shown in Figure 12, the parameter continuation method accurately captures the resonance peak of the monitored blade when the right blade is excited and its tip response is monitored. In contrast, the phase resonance method locates the resonance peak of the left (unexcited) blade, whereas the dNNMs solution falls at a frequency between the two peaks.
\begin{figure}[htbp] 
    \centering
    \begin{subfigure}{0.47\textwidth} 
        \centering 
        \includegraphics[width=\linewidth]{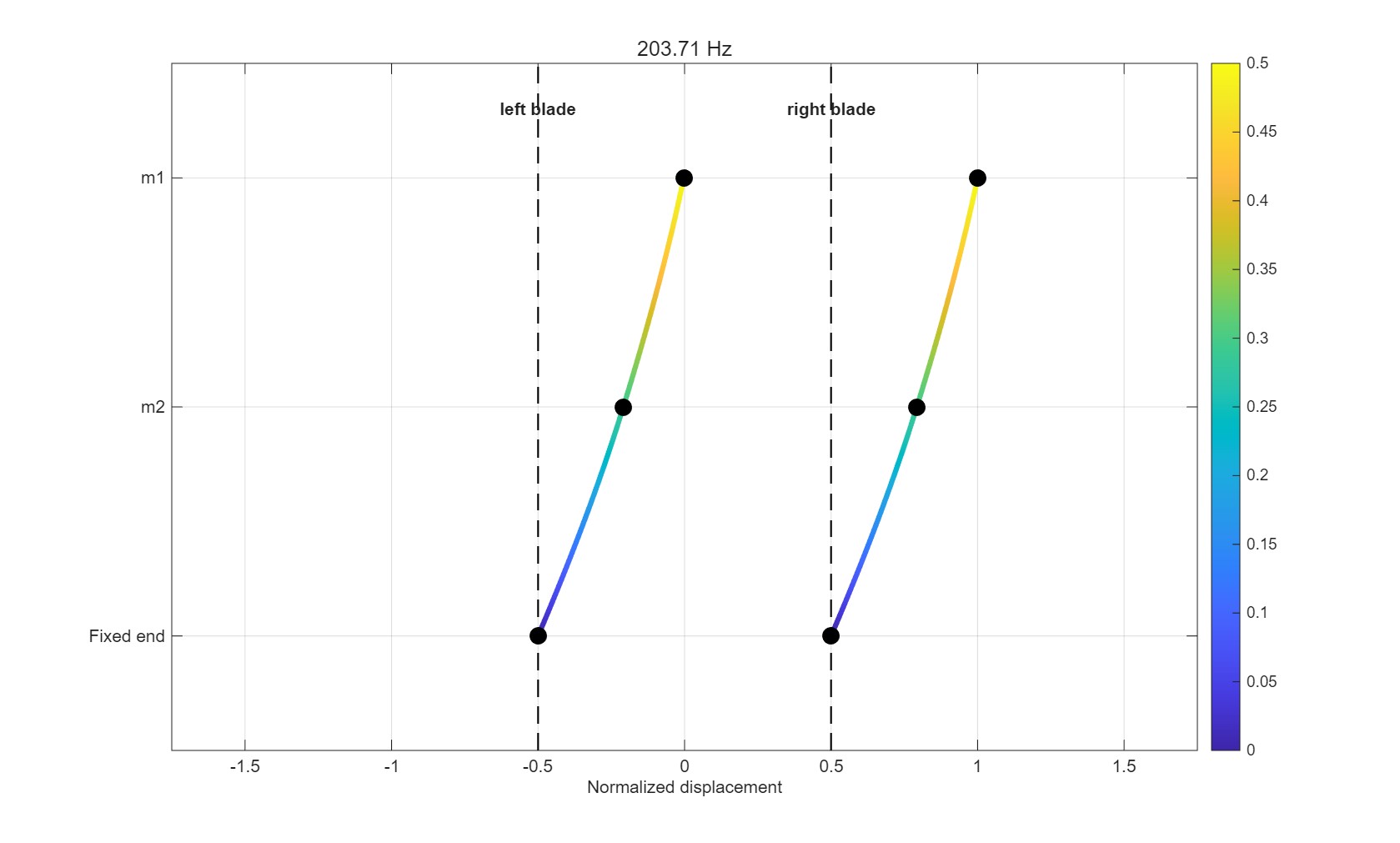}
        \caption{In-phase bending mode}
        \label{fig:sub1}
    \end{subfigure}
    \begin{subfigure}{0.47\textwidth}
        \centering
        \includegraphics[width=\linewidth]{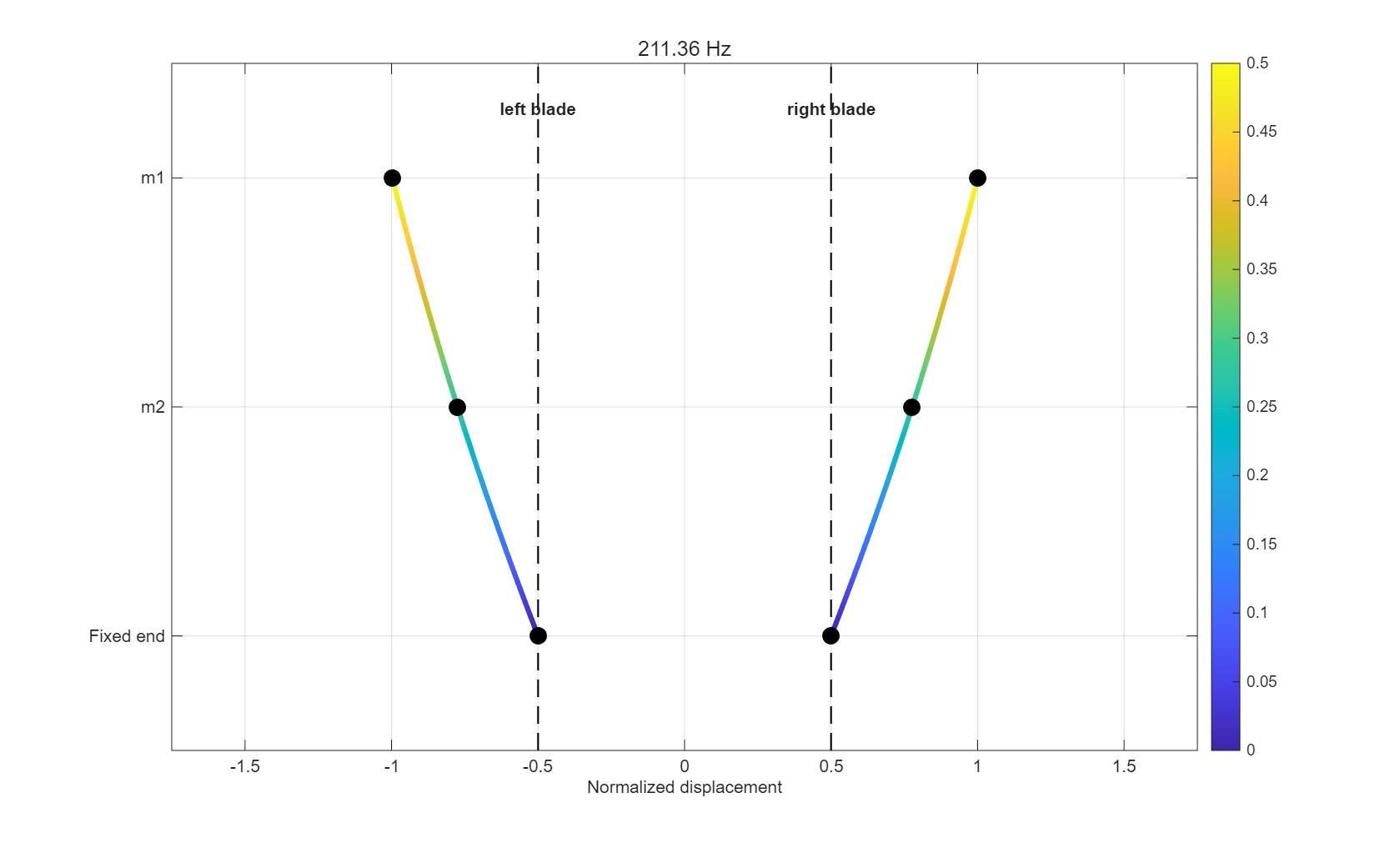}
        \caption{Out-of-phase bending mode}
        \label{fig:sub2}
    \end{subfigure}
    \caption{Linear bending vibration modes in the fully stuck contact state}
    \label{fig:overall}
\end{figure}
\begin{figure}[htbp] 
    \centering
    \begin{subfigure}{0.8\textwidth} 
        \centering 
        \includegraphics[width=\linewidth]{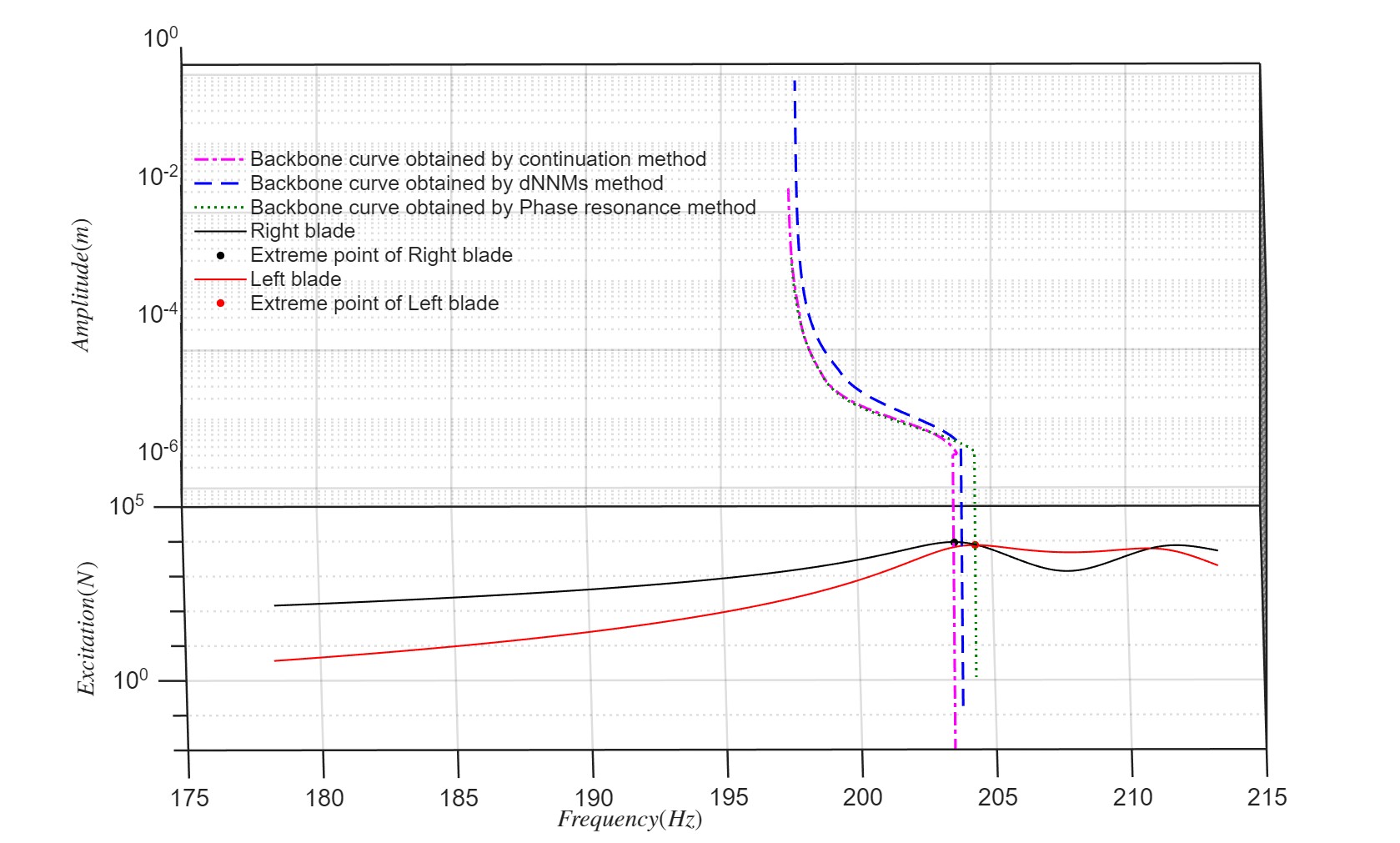}
        \caption{Three stages of the continuation process and their respective curves}
        \label{fig:sub1}
    \end{subfigure}
    \\ 
    \begin{subfigure}{0.48\textwidth}
        \centering
        \includegraphics[width=\linewidth]{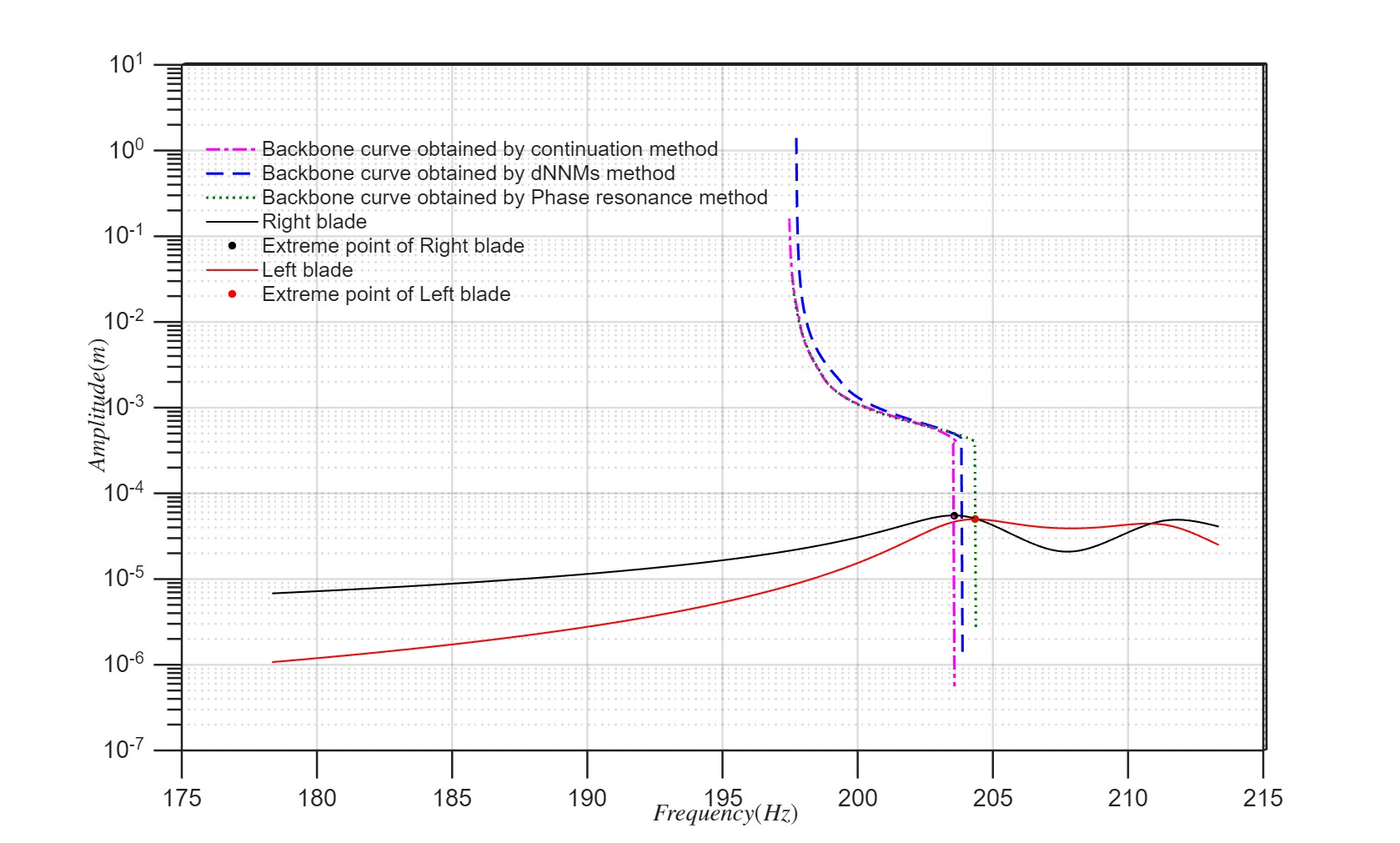}
        \caption{Frequency--amplitude plane}
        \label{fig:sub2}
    \end{subfigure}
    \hfill 
    \begin{subfigure}{0.48\textwidth}
        \centering
        \includegraphics[width=\linewidth]{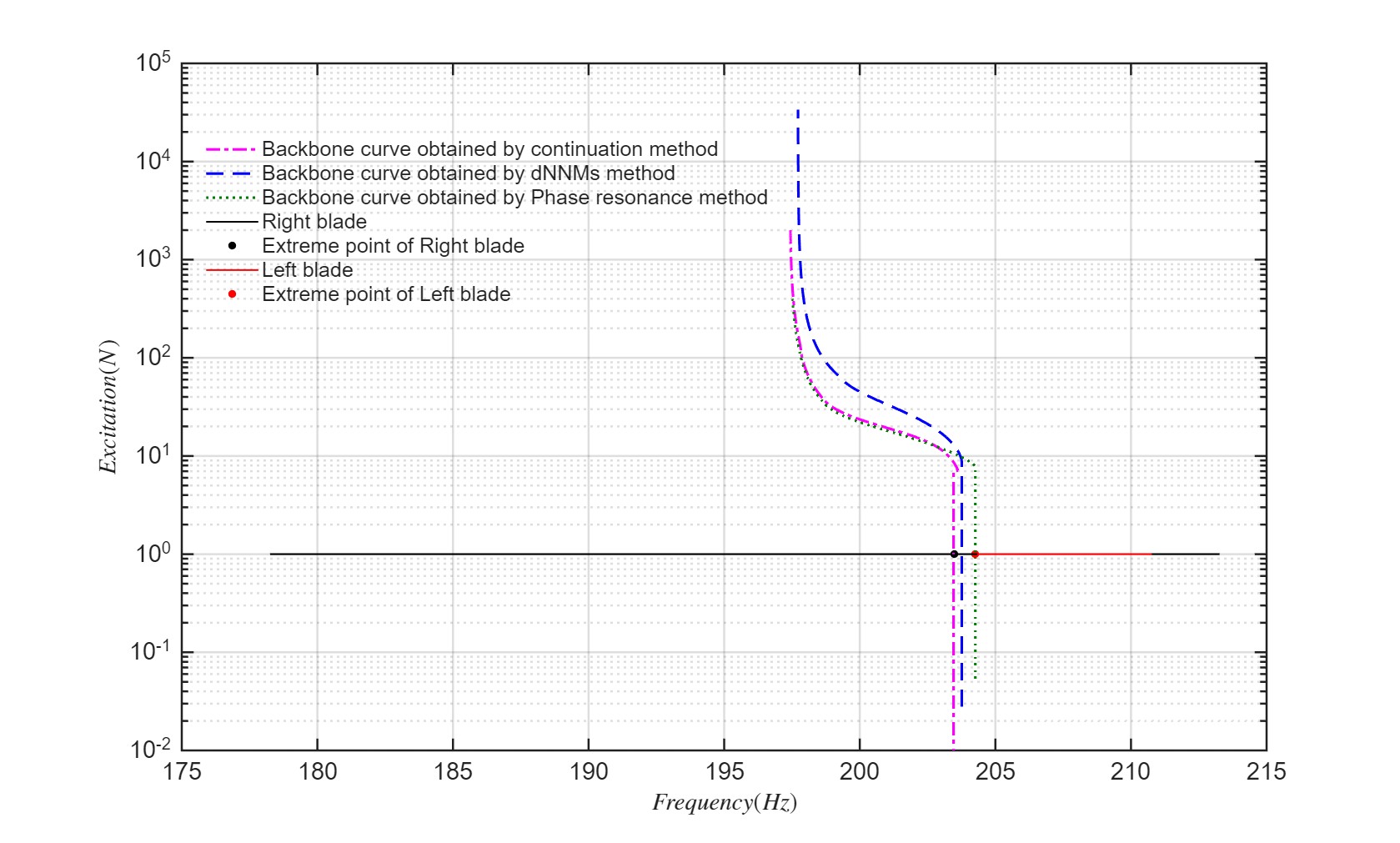}
        \caption{Frequency--excitation plane}
        \label{fig:sub3}
    \end{subfigure}
    \caption{FRCs of the left and right blade tips in the fully stuck contact state}
    \label{fig:overall}
\end{figure}

This phenomenon is mainly attributed to the large damping in the underlying linear system. As indicated by \cite{(18)cenedese2020conservative}, the backbone curve of a conservative system can be directly perturbed into the forced resonance backbone curve of its weakly damped counterpart. However, when damping is significant, the phase resonance criterion is no longer satisfactory. To verify this hypothesis, we reduced the damping ratio of the underlying linear system while maintaining a fully stuck contact state. Specifically, the stiffness-proportional damping factors $\beta$ were set to 0.000008 and 0.000003, respectively. A comparison was then made between the FRCs of the two blade tips and the backbone curves computed using the three methods.
\begin{figure}[htbp] 
    \centering
    \begin{subfigure}{0.6\textwidth} 
        \centering 
        \includegraphics[width=\linewidth]{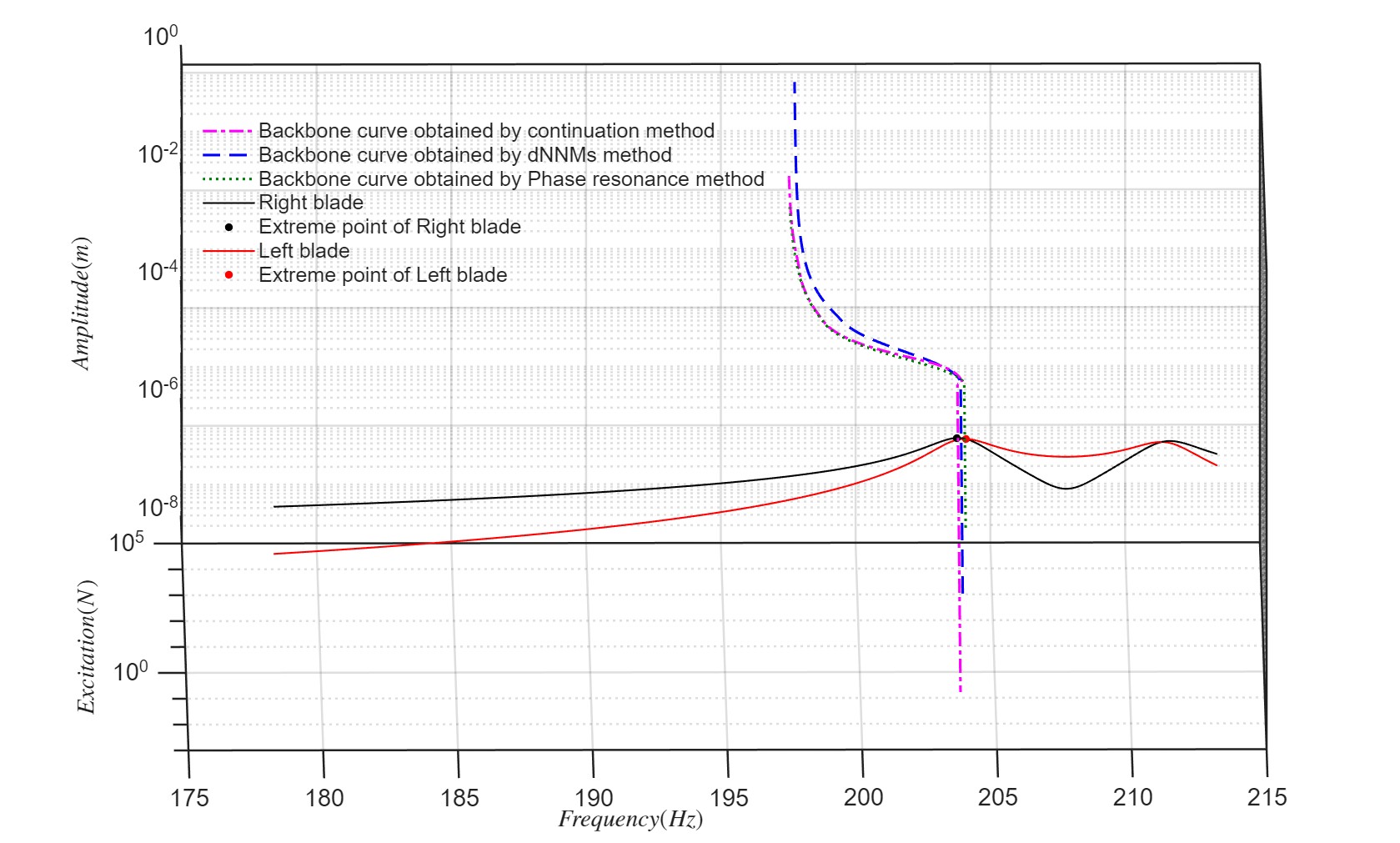}
        \caption{Damping factor $\beta = 0.000008$}
        \label{fig:sub1}
    \end{subfigure}
   \\ 
    \begin{subfigure}{0.6\textwidth}
        \centering
        \includegraphics[width=\linewidth]{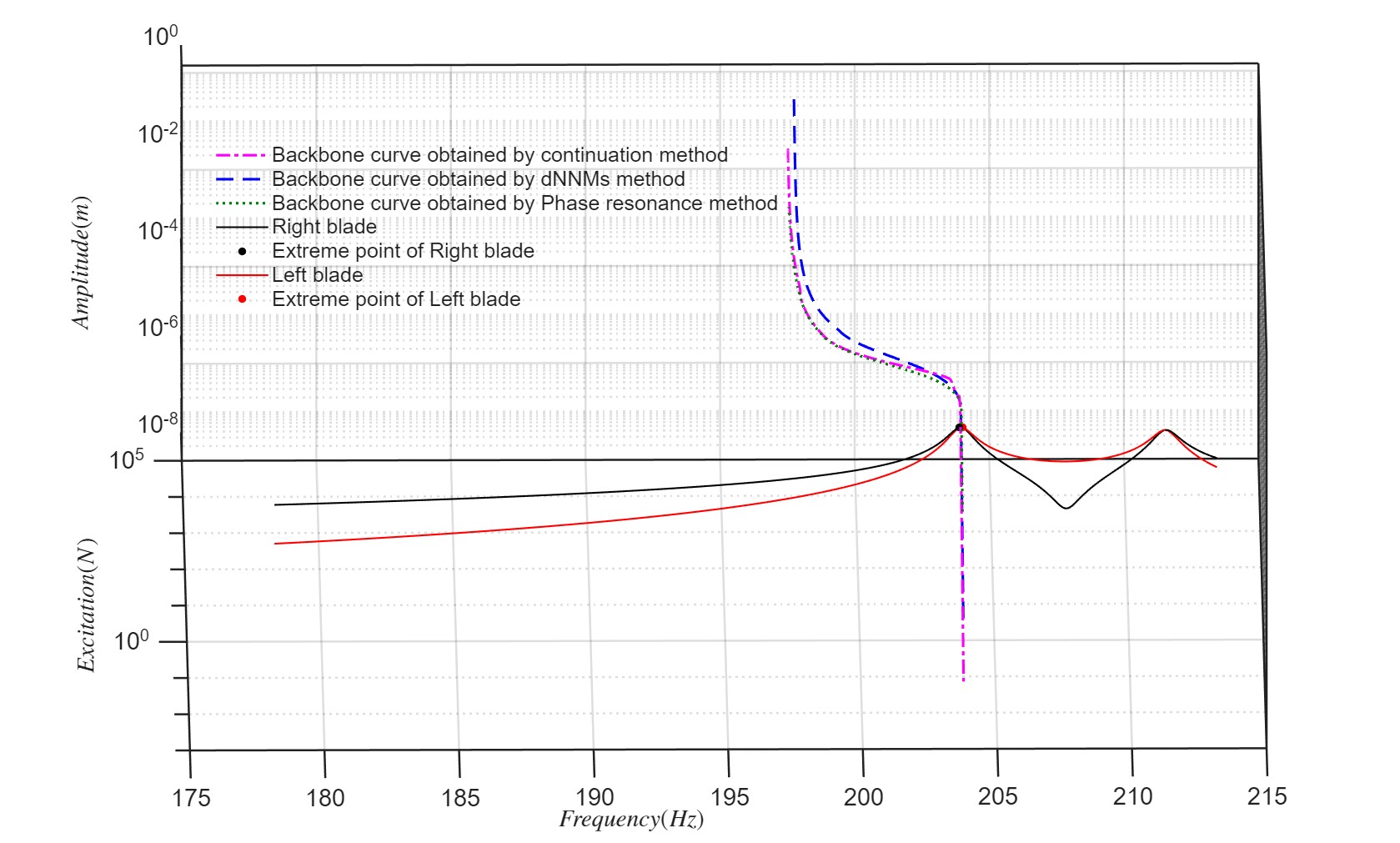}
        \caption{Damping factor $\beta = 0.000003$}
        \label{fig:sub2}
    \end{subfigure}
    \caption{Frequency response curves of the left and right blade tips in the fully stuck contact state for different damping factors}
    \label{fig:overall}
\end{figure}

In Figure 13, the resonance frequencies of the two blades tend to synchronise as the damping decreases in the stuck contact state. This observation is supported by the phase response curves of the blade-tip displacements, as shown in Figure 14. Therefore, all DOFs are concluded to reach their resonance peaks at the same frequency only in lightly damped linear or conservative nonlinear systems, thus simultaneously satisfying the phase resonance quadrature criterion. Under such conditions, computing the backbone curve using the dNNMs and phase resonance methods is considered reasonable.
\begin{figure}[htbp] 
    \centering
     \begin{subfigure}{0.47\textwidth} 
        \centering 
        \includegraphics[width=\linewidth]{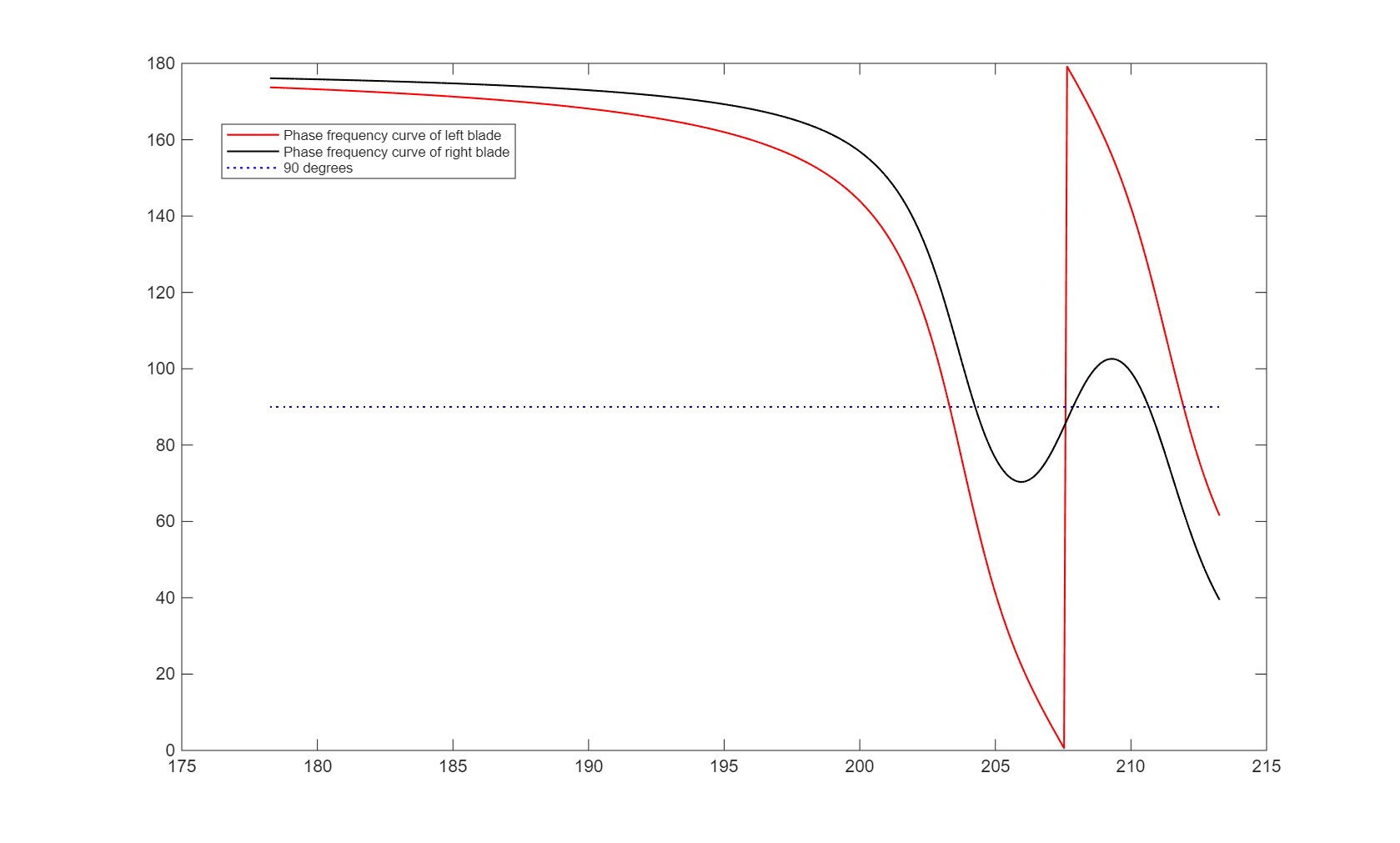}
        \caption{Damping factor $\beta = 0.000015$}
        \label{fig:sub1}
    \end{subfigure}
  \\    
    \begin{subfigure}{0.47\textwidth} 
        \centering 
        \includegraphics[width=\linewidth]{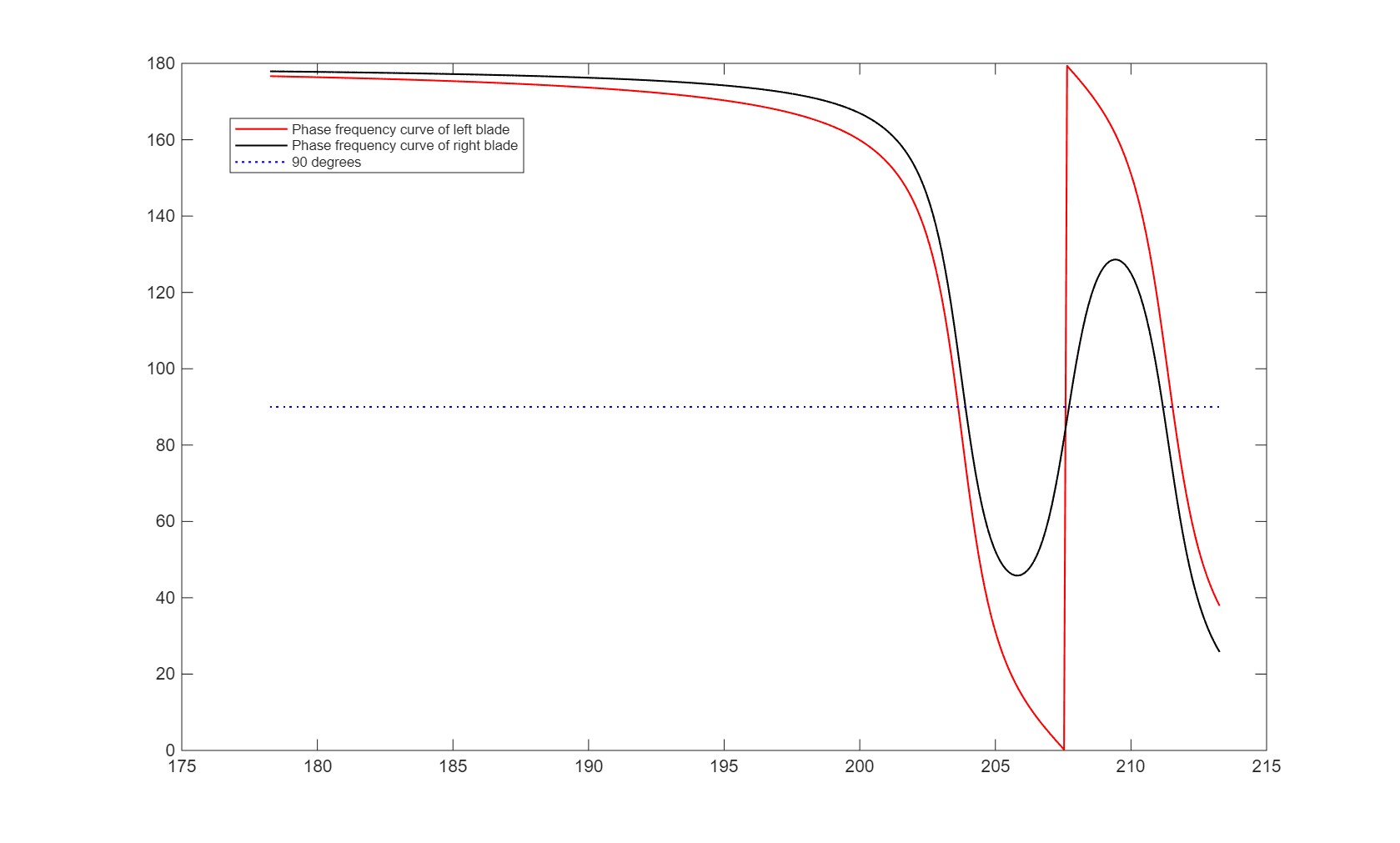}
        \caption{Damping factor $\beta = 0.000008$}
        \label{fig:sub1}
    \end{subfigure}
    \hfill 
    \begin{subfigure}{0.47\textwidth}
        \centering
        \includegraphics[width=\linewidth]{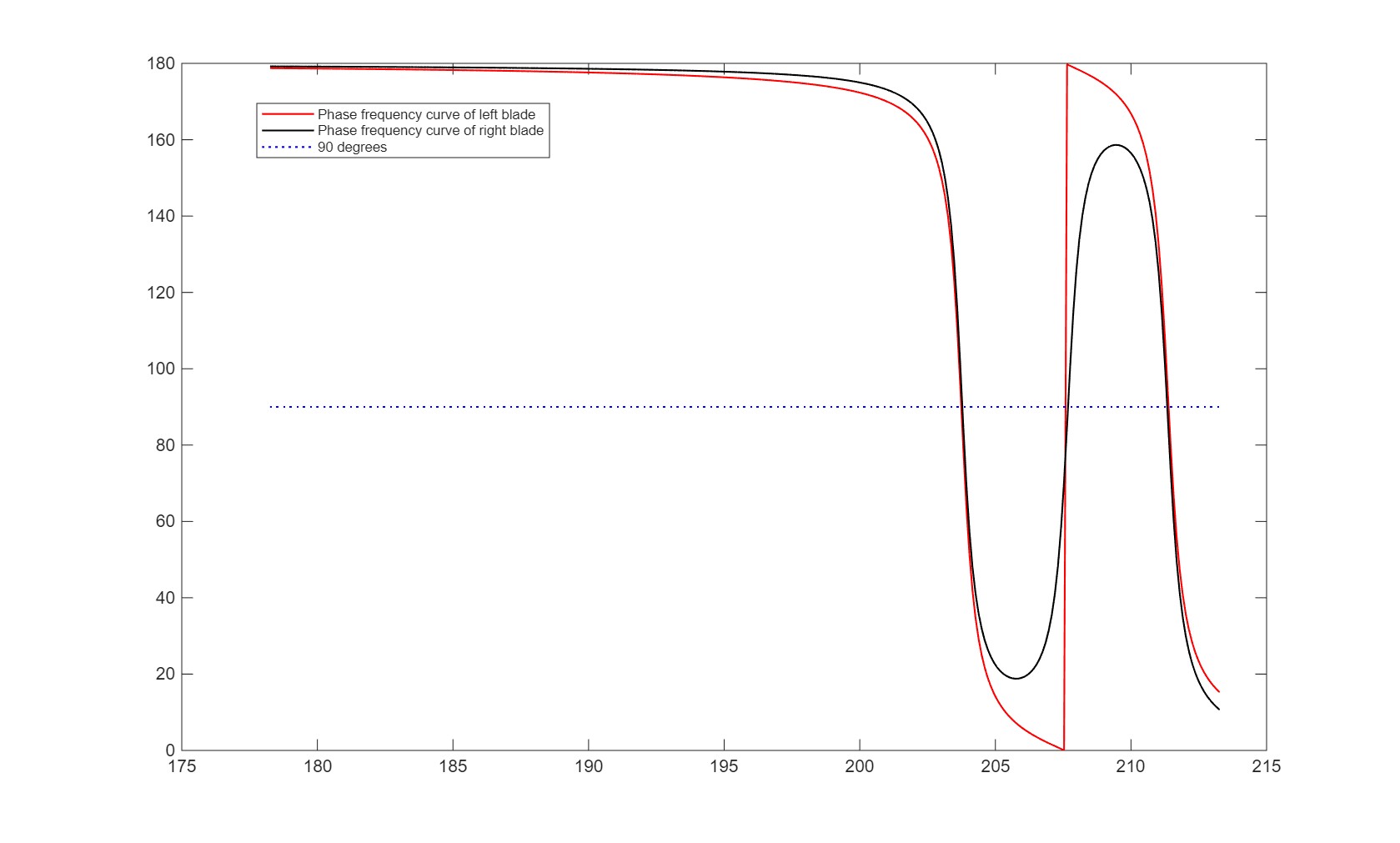}
        \caption{Damping factor $\beta = 0.000003$}
        \label{fig:sub2}
    \end{subfigure}
    \caption{Phase response curves of the left and right blade tips in the fully stuck contact state for different damping factors}
    \label{fig:overall}
\end{figure}

In addition to damping the underlying linear system, the damping introduced by the dry friction interface must be considered. We set the damping ratio of the linear system to 0.000003 and performed frequency and phase response computations under the maximum frictional damping state.
\begin{figure}[htbp] 
    \centering
     \begin{subfigure}{0.8\textwidth} 
        \centering 
        \includegraphics[width=\linewidth]{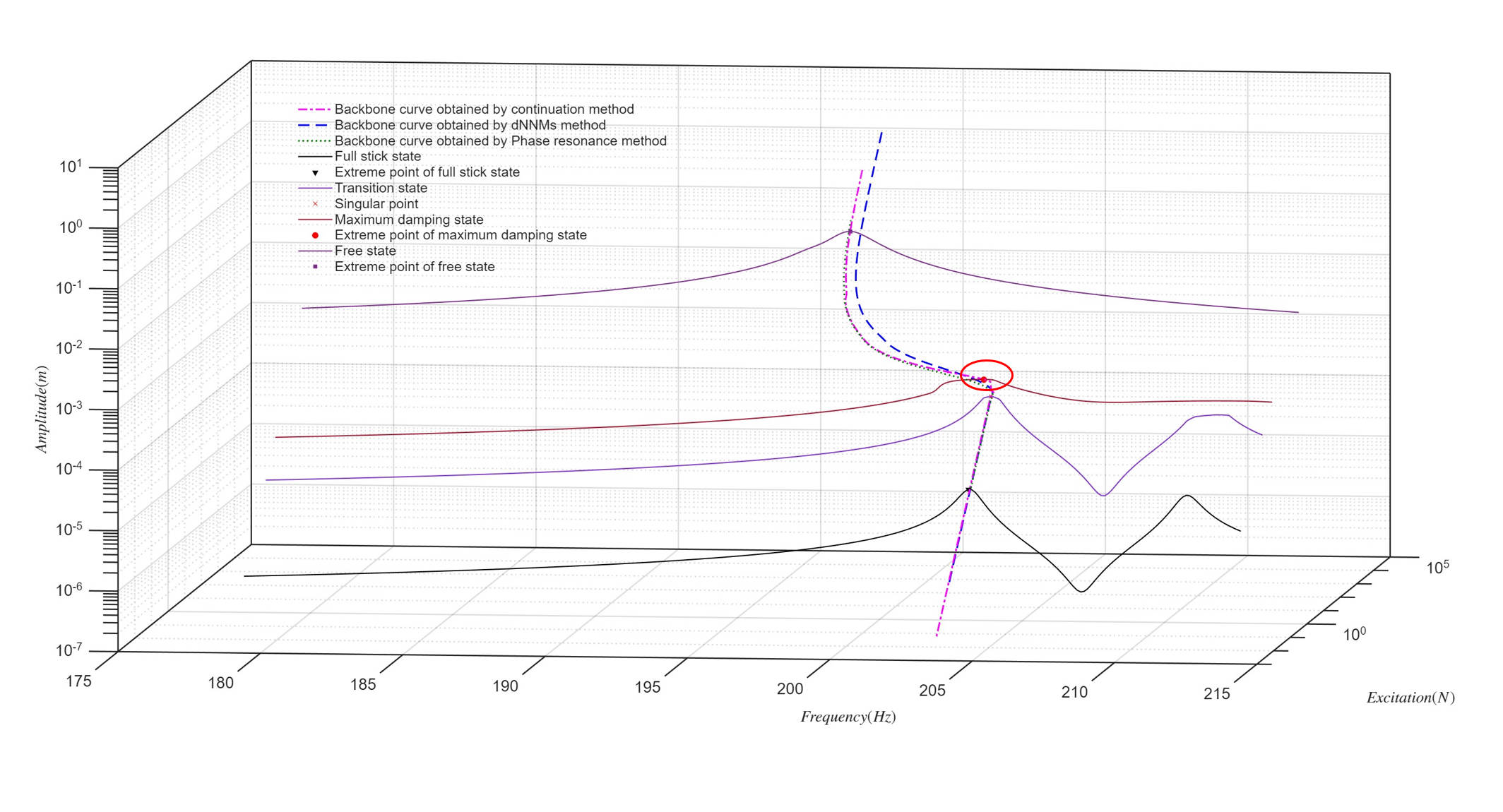}
        \caption{Damping factor $\beta = 0.000015$}
        \label{fig:sub1}
    \end{subfigure}
  \\    
    \begin{subfigure}{0.47\textwidth} 
        \centering 
        \includegraphics[width=\linewidth]{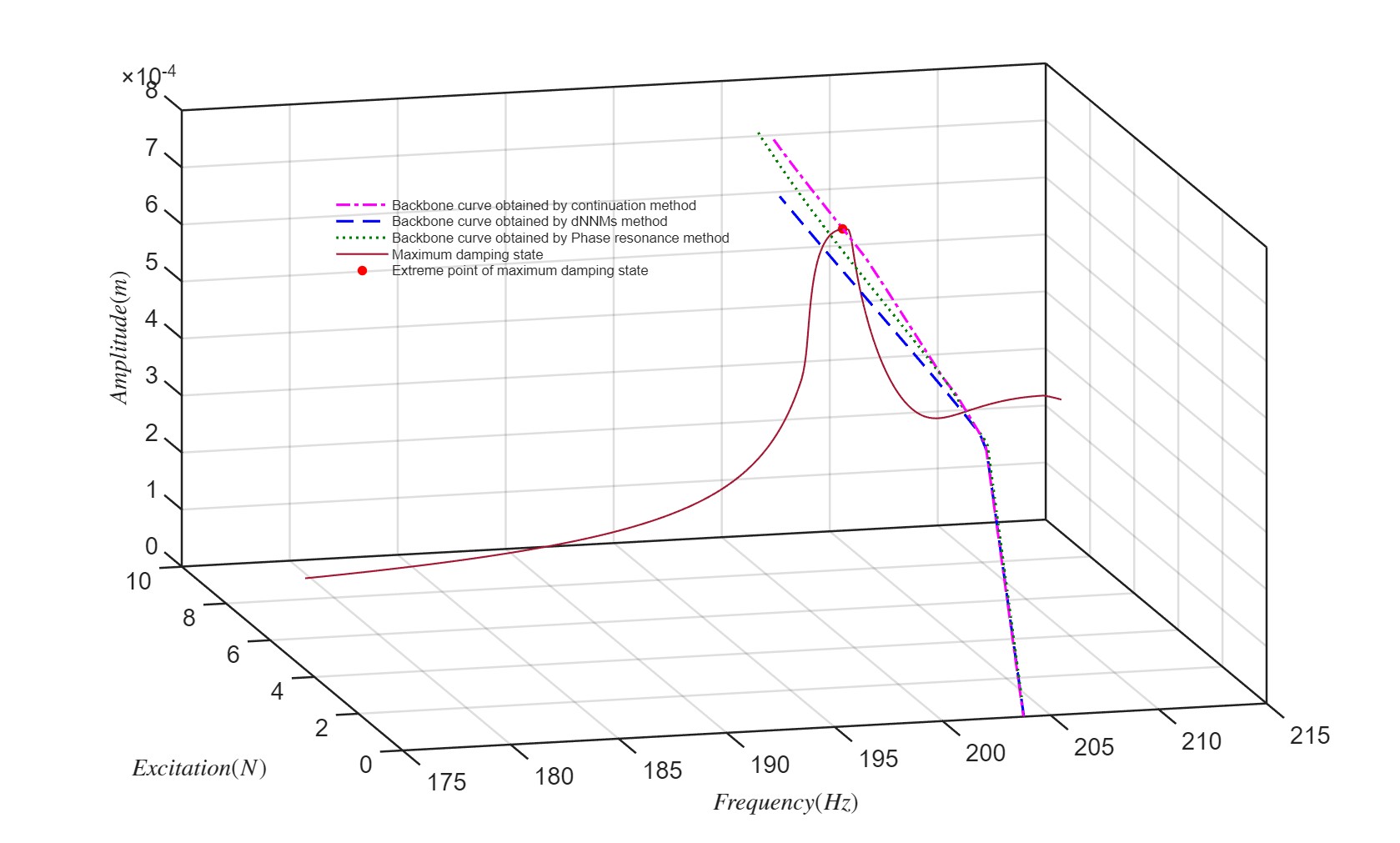}
        \caption{Damping factor $\beta = 0.000008$}
        \label{fig:sub1}
    \end{subfigure}
    \hfill 
    \begin{subfigure}{0.47\textwidth}
        \centering
        \includegraphics[width=\linewidth]{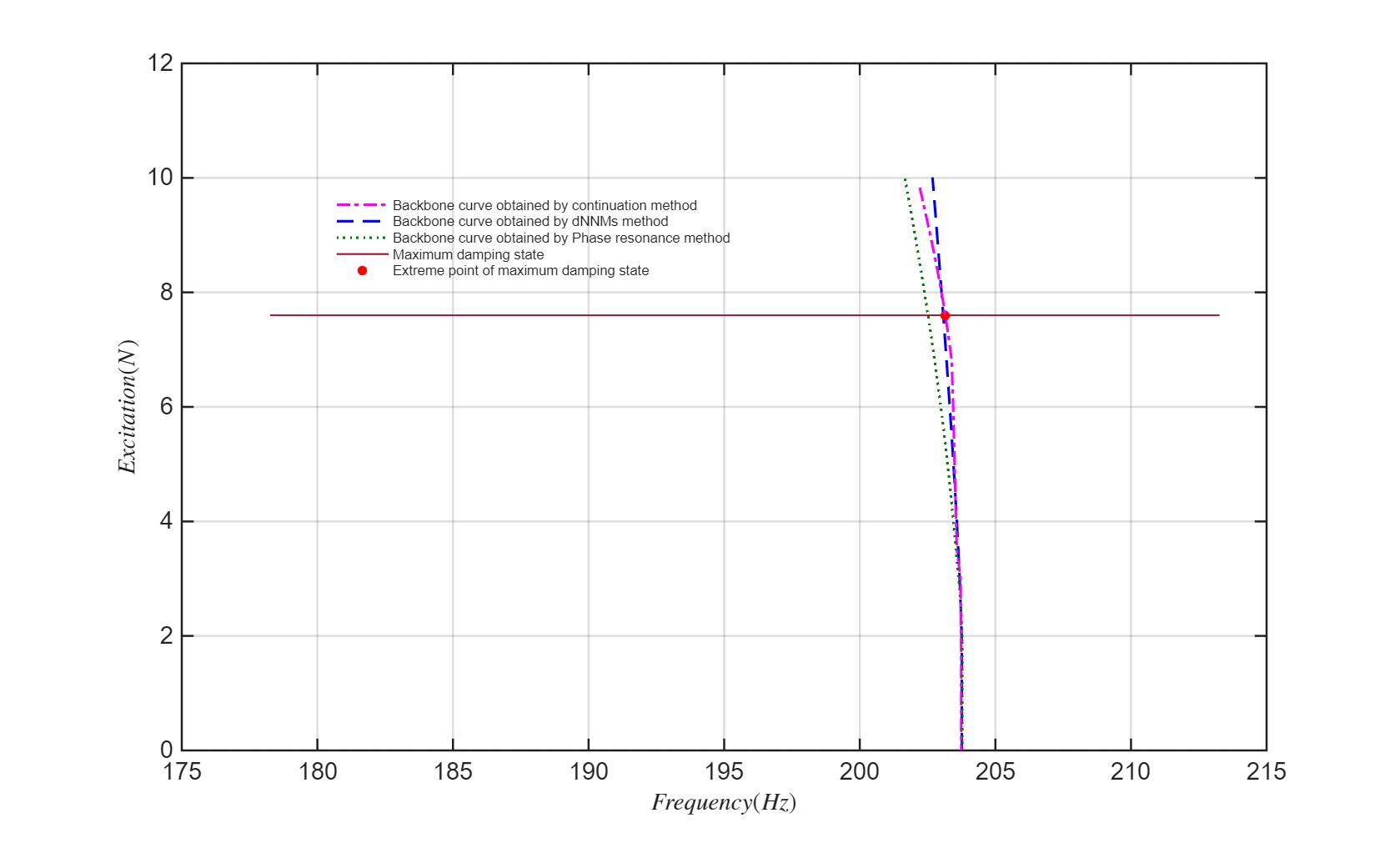}
        \caption{Damping factor $\beta = 0.000003$}
        \label{fig:sub2}
    \end{subfigure}
    \caption{Phase response curves of the left and right blade tips in the fully stuck contact state for different damping factors}
    \label{fig:overall}
\end{figure}

From the projection of the frequency response in Figure 15, it can be observed in the red-circled region that the resonance points predicted by the dNNMs and phase resonance methods deviate from the actual resonance frequency of the forced response. This deviation is primarily caused by the asynchronous vibration of the left and right blades induced by dry frictional damping. To confirm this, we compared the frequency and phase response curves of the two blade tips at this point. Under the excitation force, the asynchronous motion of the blades is significant, as shown in Figure 14. Furthermore, the dry friction nonlinearity effect is strong at this stage, leading to a noticeable nonsmooth nature of the frequency response curve.
\begin{figure}[htbp] 
    \centering
    \begin{subfigure}{0.47\textwidth} 
        \centering 
        \includegraphics[width=\linewidth]{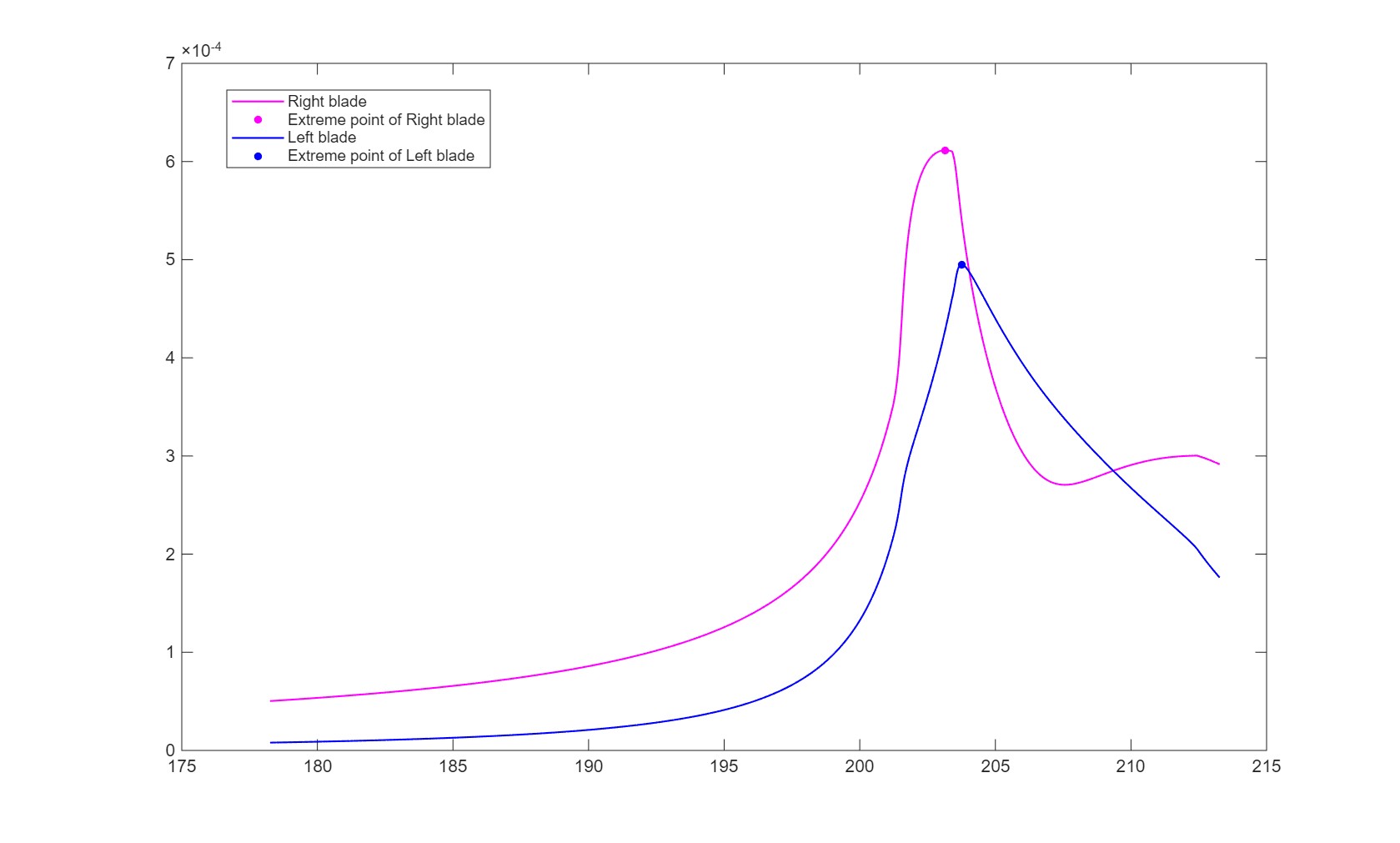}
        \caption{Amplitude response curves}
        \label{fig:sub1}
    \end{subfigure}
    \begin{subfigure}{0.47\textwidth}
        \centering
        \includegraphics[width=\linewidth]{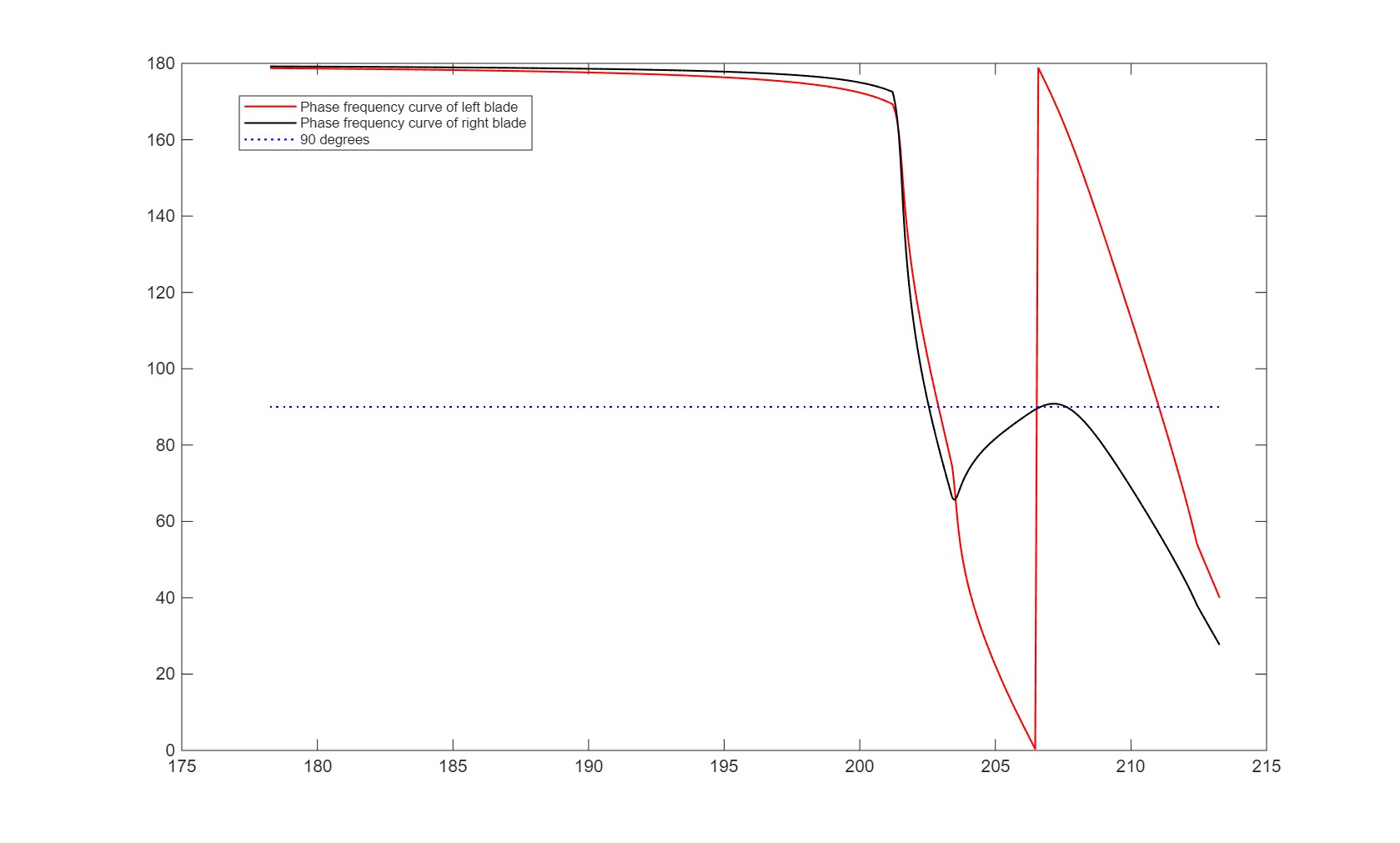}
        \caption{Phase response curves}
        \label{fig:sub2}
    \end{subfigure}
    \caption{Amplitude and phase response curves of the left and right blade tips in the maximum frictional damping state}
    \label{fig:overall}
\end{figure}

The two case studies demonstrate that the parameter continuation method exhibits superior applicability to the dNNMs and phase resonance methods, particularly under the following conditions:

\begin{enumerate}[1)]
\item When higher-order harmonics are retained in the analysis, the parameter continuation method more effectively captures their influence on the response.
\item For systems with two closely spaced modal frequencies and large underlying linear damping, the DOFs do not resonate synchronously. In this scenario, resonance identification via the dNNMs and phase resonance methods is inaccurate.
\item For systems with two closely spaced modal frequencies and small underlying linear damping, if the nonconservative nonlinearity of the system introduces significant damping effects, the dNNMs and phase resonance methods fail to accurately identify the location of the resonance peak within the region of strong nonlinear damping.
\end{enumerate}

\section{Conclusion} \label{ Conclusion}
This study proposes a novel method for accurately predicting forced resonance backbone curves in nonlinear structures with dry friction, employing an analytical Hessian Tensor of contact elements and parameter continuation techniques. The proposed method proves to be computationally more efficient than conventional frequency-stepping method. Furthermore, it demonstrates higher accuracy and broader applicability than established backbone curve computation techniques, such as damped nonlinear normal modes (dNNMs) and phase resonance methods.

The theoretical framework is established using the HBM to solve the governing equations of structures with friction nonlinearity. Based on the geometric interpretation of the forced resonance backbone curve as the response ridge on the system's resonance surface, a solution strategy was formulated using the Lagrangian multiplier method. Since the Newton iteration for this formulation requires the second derivative of the nonlinear force, the analytical Hessian Tensor for the contact element was derived from explicit expressions of contact force and stiffness. This analytical construction of the Jacobian matrix facilitates a rapid and robust numerical solution.

The accuracy and robustness of the proposed method were validated through two numerical case studies. First, the analysis of a cantilever beam with frictional ground contact demonstrated that the proposed method achieves superior convergence and accuracy compared to alternative approaches, particularly when higher-order harmonics are retained. Second, the limitations of phase-lag criterion-based methods (dNNMs and phase resonance) were highlighted using a lumped-parameter blade–damper–blade model. These conventional methods proved unreliable under two specific conditions: (1) when the underlying linear system exhibits closely spaced modes with high linear damping, and (2) when nonlinear frictional damping is significant. The proposed parameter continuation method is immune to these limitations. Consequently, it serves as a robust tool for predicting forced resonance backbone curves for general structures featuring frictional contact interfaces.
\bibliographystyle{elsarticle-num}  
\bibliography{bibtex}  
\end{document}